\documentclass[twocolumn]{aastex631}
\usepackage{tabularx}
\usepackage[utf8]{inputenc}
\usepackage{array}
\usepackage{amsmath}

\newcommand\lya{Ly$\alpha$}
\newcommand\Mearth{M$_{\oplus}$}
\newcommand\Rearth{R$_{\oplus}$}

\shorttitle{\sc AU Mic c's Atmosphere}
\shortauthors{Rockcliffe et al.}

\begin{document}

\title{\sc Far-ultraviolet flares and variability of the young M dwarf AU Mic: a non-detection of planet c in transit at Lyman-$\alpha$}

\correspondingauthor{Keighley Rockcliffe}
\email{keigh.rockcliffe@gmail.com}

\author[0000-0003-1337-723X]{Keighley E. Rockcliffe}
\affiliation{Department of Physics and Astronomy, Dartmouth College, Hanover, NH 03755, USA}

\author[0000-0003-4150-841X]{Elisabeth R. Newton}
\affiliation{Department of Physics and Astronomy, Dartmouth College, Hanover, NH 03755, USA}

\author[0000-0002-1176-3391]{Allison Youngblood}
\affiliation{NASA Goddard Space Flight Center, Greenbelt, MD 20771, USA}

\author[0000-0002-7119-2543]{Girish M. Duvvuri}
\affiliation{Department of Physics and Astronomy, Vanderbilt University, Nashville, TN 37240, USA}

\author[0000-0002-0388-8004]{Emily A. Gilbert}
\affiliation{Jet Propulsion Laboratory, California Institute of Technology, Pasadena, CA 91109, USA}

\author[0000-0002-8864-1667]{Peter Plavchan}
\affiliation{Department of Physics and Astronomy, George Mason University, Fairfax, VA 22030, USA}

\author[0000-0002-8518-9601]{Peter Gao}
\affiliation{Earth and Planets Laboratory, Carnegie Institution for Science, Washington, DC 20015, USA}

\author[0000-0001-7364-5377]{Hans-R. M\"uller}
\affiliation{Department of Physics and Astronomy, Dartmouth College, Hanover, NH 03755, USA}

\author[0000-0002-9464-8101]{Adina~D.~Feinstein}
\altaffiliation{NHFP Sagan Fellow}
\affiliation{Laboratory for Atmospheric and Space Physics, University of Colorado Boulder, UCB 600, Boulder, CO 80309}
\affiliation{Department of Physics and Astronomy, Michigan State University, East Lansing, MI 48824, USA}

\author[0000-0001-7139-2724]{Thomas Barclay}
\affiliation{NASA Goddard Space Flight Center, Greenbelt, MD 20771, USA}

\author{Eric D. Lopez}
\affiliation{NASA Goddard Space Flight Center, Greenbelt, MD 20771, USA}

\begin{abstract}

Atmospheric escape's potential to shape the exoplanet population motivates detailed observations of systems actively undergoing escape. AU Mic is a young and active M dwarf hosting two close-in transiting sub- to Neptune-sized planets. Atmospheric escape was previously detected on the inner planet b, with radially-blown neutral hydrogen producing $\sim30$\% blue-shifted absorption in Lyman-$\alpha$. We obtained one {\it HST}/STIS transit of the outer planet c, to search for the planet's escaping atmosphere in transmission at Lyman-$\alpha$ and compare with AU Mic b. We detected $6$ short-duration flares in \ion{Si}{4} and \ion{C}{4}, of which only one corresponded to a Lyman-$\alpha$ flare. We identified longer-duration stellar variability at the tens of percent level for lines less sensitive to stellar activity, including \ion{O}{1}, \ion{C}{2} and Lyman-$\alpha$, which inhibits detection of an exosphere. We do not report absorption associated with an exosphere containing neutral hydrogen or any metals detectable in the far-ultraviolet, and discuss the implications of the non-detection. This work highlights the importance of 1) careful consideration of stellar variability in atmospheric escape observations, and 2) the dual-influence of photoionization and stellar wind when interpreting and modeling atmospheric escape.

\end{abstract}

\section{Introduction} \label{sec:intro}

The ever-growing number of detected exoplanets illuminates trends in exoplanet demographics that may be artifacts from planet formation and evolution. Atmospheric escape is an evolutionary mechanism thought to have significantly shaped the current population of exoplanets, leaving imprints like the ``hot Neptune desert" \citep{2016NatCo...711201L} and the ``radius gap" \citep{2016ApJ...831..180C,2017AJ....154..109F,2017ApJ...847...29O}.

Theory predicts that short-period planets born with large gaseous envelopes --- Neptune to Saturn-sized --- will eventually lose most to all of their atmospheres due to heating from intense high-energy radiation \citep[e.g.,][]{2013ApJ...776....2L,2014ApJ...795...65J,2018MNRAS.479.5012O}. Photoevaporation is the process in which incident EUV and X-ray (XUV) radiation deposits heat into an atmosphere and causes gas to escape the planet's gravitational potential. While alternative explanations exist for the observed dearth of close-in planets with thick atmospheres, such as core-powered mass loss \citep{2018MNRAS.476..759G} and primordial formation mechanisms \citep{2021ApJ...908...32L}, photoevaporation remains a strong influence on low density planets in extreme XUV environments.

\subsection{Observations of atmospheric escape}

The first observations of atmospheric escape were made two decades ago for short-period, Jupiter-sized planets (hot Jupiters) like HD 209458 b \citep{2003Natur.422..143V}. The host star's \ion{H}{1} Lyman-$\alpha$ emission line (\lya; 1215.67 \AA) was used to observe the planet's escaping neutral hydrogen in transmission; this measurement is possible because of neutral hydrogen's large photoabsorption cross section at 1215.67 \AA.

The landmark {\it Hubble Space Telescope (HST)}/Space Telescope Imaging Spectrograph \citep[STIS;][]{1998PASP..110.1183W} observation of the hot Neptune Gl 436 b provided the first detection of atmospheric escape off of a planet smaller than Jupiter \citep{2014ApJ...786..132K,2015Natur.522..459E,2017A&A...605L...7L}. Atmospheric escape, whether photoevaporation or core-powered mass loss, is predicted to be most influential during the first 100 - 1000 Myr of an exoplanet's lifetime \citep{2018MNRAS.478.1193K}. However, Gl 436 b, and other hot Neptunes with atmospheric-stripping levels of escape (e.g., GJ 3470 b, \citealt{2018A&A...620A.147B}; HAT-P-11 b, \citealt{2022NatAs...6..141B}) are older than $1$ Gyr. This brings into question whether they are actually representative of the broader evolutionary trend of most hot Neptunes, or if their orbital evolution led to photoevaporation at older ages. Gl 436 b, GJ 3470 b, and HAT-P-11 b show evidence that they may have migrated closer to their hosts later in life and therefore only recently started experiencing atmospheric escape, potentially explaining why catastrophic escape is detected when the systems are $> 1$ Gyr old \citep{2022A&A...663A.160B,2022ApJ...931L..15S,2023A&A...669A..63B}.

The search for atmospheric escape detections on young to intermediate aged planets ($< 1$ Gyr old) began with K2-25 b. \cite{2021AJ....162..116R} used the same \lya\ transmission technique to search for neutral hydrogen escaping off of this $\sim 700$ Myr planet, to no avail. They postulate that the large amount of ionizing radiation from the young star is ionizing the escaping material too quickly to be observed in transmission at \lya. It is also possible that interactions with the stellar wind could confine the exosphere to a smaller, undetectable area, or the planet could be denser than originally thought. These hypotheses could also explain the subsequent non-detection of atmospheric escape on the $400$ Myr hot Neptune HD 63433 b \citep{2022AJ....163...68Z}. However, \cite{2022AJ....163...68Z} were able to detect atmospheric escape around HD 63433 c, which is exterior to b. \cite{feinstein21} and \cite{schlawin21} found tentative evidence of H$\alpha$ exospheres from the $30-40$ Myr planet V1298~Tau~c, which orbits at a similar to distance as HD 63433 b, but results were inconclusive due to H$\alpha$ variability from the host star. \cite{2024RNAAS...8...86F} were also unable to detect metal escape off of V1298~Tau~c. These observations show that there is no obvious trend between orbital separation and whether a atmospheric escape is detected or not in \lya, even though the effects of atmospheric photoionization and stellar winds decrease radially from the host star.

\subsection{The AU Mic planetary system}

In this work, we continue our study of atmospheric escape in the AU Mic system. AU Mic is a $23$ Myr pre-main sequence star within the $\beta$ Pic moving group \citep{2004ARA&A..42..685Z,2014MNRAS.445.2169M}. It has a debris disk and three confirmed planets: AU Mic b and c and, likely orbiting in-between them, a third non-transiting planet AU Mic d \citep{2009ApJ...698.1068P, 2020Natur.582..497P, 2021AA...649A.177M, 2022AJ....163..147G, 2023AJ....166..232W}.

AU Mic b is larger than Neptune, orbits closer to its host with a period of $8.463$ days, and has several possible masses between $10.2 - 20.12$ \Mearth\ \citep{2021AJ....162..295C,2021MNRAS.502..188K,2022MNRAS.512.3060Z,2023MNRAS.525..455D}. \cite{hirano20} did not detect the atmosphere of AU Mic b in \ion{He}{1} transmission, and \cite{2024RNAAS...8...86F} did not detect its atmosphere in metal species transmission. \cite{2023AJ....166...77R} detected AU Mic b's neutral hydrogen exosphere, finding a $\sim 30$\% transit depth at \lya\ in one observation and a non-detection in a second observation. The variability in detection may be due to the changing ionization state of the exosphere, interactions with the stellar wind causing various exosphere shapes, or a combination of both. \cite{2023AJ....166...77R} did not find evidence that AU Mic's far-ultraviolet variability impacted the detection of AU Mic b at \lya.

\begin{figure*}[ht!]
\epsscale{1.15}
\plotone{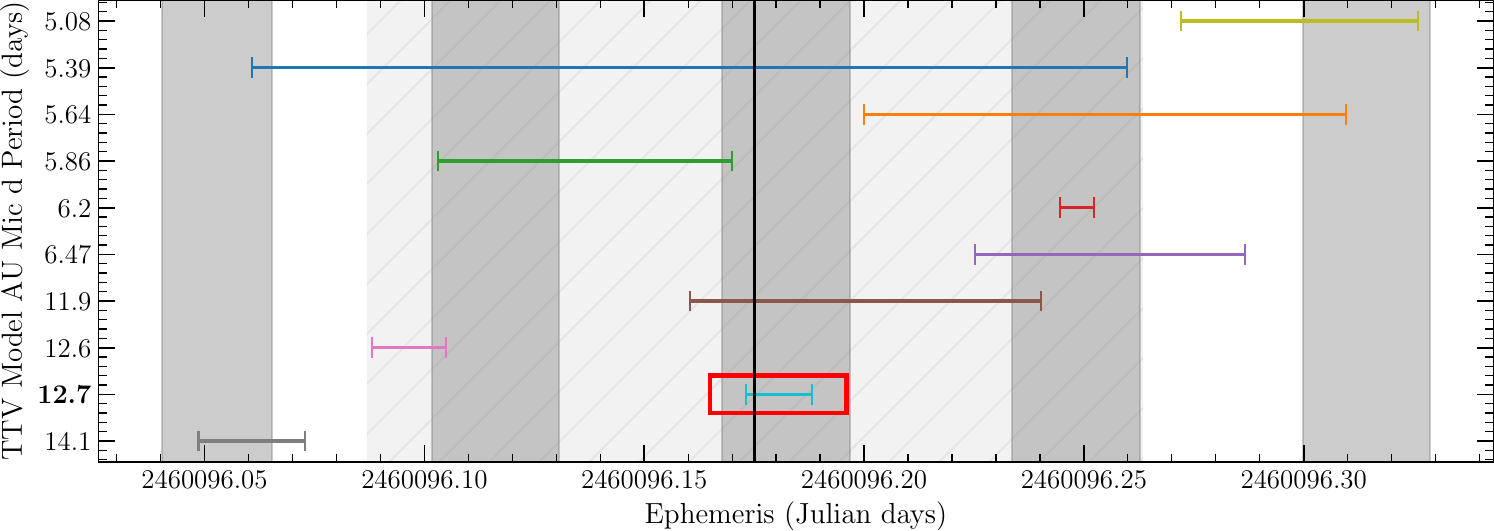}
\caption{All potential TTV model periods of AU Mic d (y-axis labels) and their associated impact on the mid-transit ephemeris of AU Mic c (horizontal lines of different colors) from \cite{2023AJ....166..232W} around our data acquisition. The error bars represent the uncertainty in the ephemeris predicted by that model. The most favorable model, $12.7$ days, is within the red rectangle. The gray regions indicate the {\it HST} exposures presented in this work. The black vertical line is the AU Mic c ephemeris used for this work, which was calculated using the planet's epoch and period reported in \cite{2023AJ....166..232W}, which was recently confirmed by \cite{2025A&A...694A.137B}. The light hatched region is the $4.236$ hour white-light transit duration assuming the black vertical line as mid-transit. \label{fig:ephem}}
\end{figure*}

AU Mic c is a little smaller than Neptune and has an orbital period of $18.859$ days (see Table~\ref{tab:prop} for system properties). \cite{2023AJ....166..232W} confirmed the presence of AU Mic d using transit-timing variations (TTVs), and subsequently modeled ephemerides for all the AU Mic planets over the next few years. The AU Mic c ephemerides strongly depend on the correct period for AU Mic d. All possible AU Mic c ephemerides (corresponding to different potential AU Mic d periods) from \cite{2023AJ....166..232W} are illustrated in Figure~\ref{fig:ephem}. The data used in this work span the entire range of uncertainty in AU Mic c's ephemeris. The statistically and dynamically favored TTV model (P$_{\text{d}}=12.7$ days) lies within $\sim10$ minutes of the ephemeris used to analyze our observations. While this paper was under review, \cite{2025A&A...694A.137B} corroborated a period of 12.6 day for AU Mic d and our assumed ephemeris for AU Mic c's transit. It is possible, however, that the transit midpoint falls near the beginning or end of our observing window, such that we only observed half the transit. In either of those cases, based on previous \lya\ transits of hydrogen exospheres which span from about an hour prior to a few hours passed white-light mid-transit, AU Mic c's exosphere could still be detectable if present \citep[e.g.,][]{2015Natur.522..459E,2018A&A...620A.147B,2022AJ....163...68Z}.

A search for atmospheric escape on AU Mic c and comparison to AU Mic b could, by leveraging the power of keeping many system properties the same, provide further constraints on how escape varies with orbital separation and planet properties. In Section~\ref{sec:data}, we present {\it HST}/STIS far-ultraviolet observations of AU Mic c in transit. \lya\ and metal line light curves are shown and investigated for variability in Section~\ref{sec:lcurve}. We characterize AU Mic c's high-energy environment and mass loss rate in Section~\ref{sec:highen}. Section~\ref{sec:discuss} concludes with a summary of our analysis and discussion of implications.

\begin{deluxetable*}{lccl}[t!]
\tablecaption{AU Mic and AU Mic c properties. \label{tab:prop}}
\tablecolumns{4}
\tablewidth{0pt}
\tablehead{
\colhead{Properties (Symbol)} &
\colhead{Value} &
\colhead{Units} &
\colhead{Reference}
}
\startdata
Earth-system distance (d) & $9.722100 \pm 0.0004625$ & pc & \cite{2018AA...616A...1G} \\
Age ($\tau$) & $22 \pm 3$ & Myr & \cite{2014MNRAS.445.2169M} \\
Right ascension ($\alpha$) & 20:45:09.87 & hh:mm:ss & ... \\
Declination ($\delta$) & $-$31:20:32.82 & dd:mm:ss & ... \\
Spectral type & M1Ve & & \cite{2015arXiv151001731T} \\
Bolometric luminosity (L$_{\text{bol}}$) & $-1.038 \pm0.049$ & $\log_{10}$L$_{\odot}$ & \cite{2023AJ....166..232W}\\
Stellar mass (M$_{\star}$) & $0.510^{+0.028}_{-0.027}$ & M$_{\odot}$ & \cite{2023AJ....166..232W} \\
Stellar radius (R$_{\star}$) & $0.744^{+0.023}_{-0.021}$ & R$_{\odot}$ & \cite{2023AJ....166..232W} \\
Stellar rotation period (P$_{\star}$) & $4.856 \pm0.003$ & days & \cite{2023MNRAS.525..455D} \\
Radial velocity (v$_{\star}$) & $8.7 \pm0.2$ & km s$^{-1}$ & \cite{2017AA...603A..54L} \\
Epoch (t$_0$) & $2458342.2240^{+ 0.0019}_{- 0.0018}$ & BJD & \cite{2023AJ....166..232W} \\
Transit duration & $4.236 \pm0.029$ & hours & \cite{2023AJ....166..232W} \\
Planetary mass measurements (M$_{\text{p}}$) & $22.2 \pm6.7$ & \Mearth & \cite{2022MNRAS.512.3060Z} \\
... & $14.5 \pm3.4$ & \Mearth & \cite{2024AA...689A.132M} \\
... & $14.2^{+4.8}_{-3.5}$ & \Mearth & \cite{2023MNRAS.525..455D} \\
... & $13.6 \pm11.4$ & \Mearth & \cite{2021AA...649A.177M} \\
... & $9.600^{+2.070}_{-2.310}$ & \Mearth & \cite{2021AJ....162..295C} \\
Planetary radius (R$_{\text{p}}$) & $2.52 \pm0.25$ & \Rearth & \cite{2023AJ....166..232W} \\
Orbital period (P$_{\text{p}}$) & $18.85969 \pm0.00008$ & days & \cite{2023AJ....166..232W} \\
Semi-major axis (a) & $0.1108 \pm0.0020$ & AU & \cite{2023AJ....166..232W} \\
\enddata
\end{deluxetable*}

\section{Far-Ultraviolet Observations} \label{sec:data}

\subsection{New observations of one transit of AU Mic c} \label{sec:newdata}

One transit of AU Mic c was observed with {\it HST}/STIS on 31 May 2023 (HST-GO-17219; PI: Rockcliffe); a second transit was unsuccessful due to incorrect position information (Table~\ref{tab:obs}). {\it HST}'s frequent South Atlantic Anomaly (SAA) crossings limit visits using STIS's MAMA detectors to five consecutive orbits. The need to avoid SAA crossings, frequent gyro issues, and the time critical nature of transits make scheduling such experiments challenging for {\it HST}.

\begin{deluxetable*}{lcccccl}[htbp]
\tablecaption{Observation log for all visits to AU Mic with {\it HST}/STIS by the start of this work. We include whether the observation was a success (S) or failure (F). All successful observations were used in this work. \label{tab:obs}}
\tablecolumns{7}
\tablewidth{0pt}
\tablehead{
\colhead{Visit} &
\colhead{Transiting Planet} &
\colhead{Dataset} &
\colhead{S/F} &
\colhead{Start Time (BJD)} &
\colhead{Exposure Time (s)} &
\colhead{DOI}
}
\startdata
September 1998 & None & {\tt o4z301010} & S & 1998-09-06 12:17:14 & 2130.179625 & 10.17909/ekgk-8d20 \\
... & ... & {\tt o4z301020} & S & 1998-09-06 13:40:38 & 2660.188625 & ... \\
... & ... & {\tt o4z301030} & S & 1998-09-06 15:17:23 & 2660.189375 & ... \\
... & ... & {\tt o4z301040} & S & 1998-09-06 16:54:08 & 2655.1825 & ... \\
July 2020 & b & {\tt oe4h01010} & S & 2020-07-02 00:35:25 & 2306.173 & 10.17909/5hz4-vx12 \\
... & ... & {\tt oe4h01020} & S & 2020-07-02 02:04:56 & 2848.183125 & ... \\
... & ... & {\tt oe4h02010} & S & 2020-07-02 05:26:47 & 2306.188375 & ... \\
... & ... & {\tt oe4h02020} & S & 2020-07-02 06:51:06 & 2848.191875 & ... \\
... & ... & {\tt oe4h02030} & S & 2020-07-02 08:26:29 & 2848.16875 & ... \\
... & ... & {\tt oe4h03010} & S & 2020-07-03 00:26:20 & 2306.16775 & ... \\
October 2021 & b & {\tt oe4h04010} & S & 2021-10-19 02:31:08 & 2416.17925 & ... \\
... & ... & {\tt oe4h04020} & S & 2021-10-19 03:59:43 & 2758.128875 & ... \\
... & ... & {\tt oe4h04030} & S & 2021-10-19 07:10:17 & 2758.178875 & ... \\
... & ... & {\tt oe4h04040} & S & 2021-10-19 05:34:59 & 2758.1895 & ... \\
... & ... & {\tt oe4h04050} & S & 2021-10-19 08:45:33 & 2758.1945 & ... \\
... & ... & {\tt oe4h05010} & S & 2021-10-19 10:27:31 & 2416.186125 & ... \\
November 2022 & c & {\tt oexq01010} & F & 2022-11-04 07:26:24 & 2406.1105 & None \\
... & ... & {\tt oexq01020} & F & 2022-11-04 08:54:54 & 2758.12375 & ... \\
... & ... & {\tt oexq02010} & F & 2022-11-05 02:28:41 & 2406.19325 & ... \\
... & ... & {\tt oexq02020} & F & 2022-11-05 03:57:10 & 2758.187625 & ... \\
... & ... & {\tt oexq02030} & F & 2022-11-05 05:32:22 & 2758.171125 & ... \\
... & ... & {\tt oexq02040} & F & 2022-11-05 07:07:33 & 2758.189625 & ... \\
... & ... & {\tt oexq02050} & F & 2022-11-05 08:42:45 & 2758.191125 & ... \\
May 2023 & c & {\tt oexq03010} & F & 2023-05-30 16:22:07 & 2153.173625 & None \\
... & ... & {\tt oexq03020} & F & 2023-05-30 17:50:29 & 2505.0685 & None \\
... & ... & {\tt oexq04010} & S & 2023-05-31 12:58:05 & 2153.19325 & doi:10.17909/qz7b-bf83 \\
... & ... & {\tt oexq04020} & S & 2023-05-31 14:26:28 & 2505.174125 & ... \\
... & ... & {\tt oexq04030} & S & 2023-05-31 16:01:32 & 2505.1945 & ... \\
... & ... & {\tt oexq04040} & S & 2023-05-31 17:36:36 & 2505.189375 & ... \\
... & ... & {\tt oexq04050} & S & 2023-05-31 19:11:41 & 2505.168875 & ... \\
October 2023 & None & {\tt oexq53010} & F & 2023-10-09 18:36:00 & 0 & None \\
... & ... & {\tt oexq53020} & F & 2023-10-09 20:04:23 & 2505.1915 & None \\
\enddata
\end{deluxetable*}

Five successful $\sim 2500$s exposures were obtained spanning the white-light transit with STIS in its FUV-MAMA mode. We used an aperture of $0.2$\arcsec~x $0.2$\arcsec\ and the E140M echelle grating. The resulting exposures covered $1144-1729$ \AA\ at R $\sim 45800$. Each science exposure was taken while in TIME-TAG mode, which records the time of arrival of each photon observed by the instrument. We split each exposure using {\tt inttag} into ten sub-exposures ($\sim 250$s) before reduction and extraction. We used {\tt stistools.x1d.x1d} to reduce and extract the spectra from each of the sub-exposures. The {\tt calstis} pipeline within {\tt stistools} automatically found the best slit location for extraction.

Two out-of-transit orbits that took place $\sim19$ hours before the data presented in this work failed (see Table~\ref{tab:obs}). {\it HST}'s Fine Guidance Sensors failed to acquire the guide stars and subsequently failed to acquire the target. Repeat observations of these two orbits experienced the same Fine Guidance Sensor failure. Given the challenges that FUV transits with {\it HST} face and the need to use {\it HST} judiciously, we elected not to request a repeat of the entire visit.

The uncertainty calculation within {\tt stistools} does not accurately estimate uncertainties for small photon counts, which is applicable to the shorter sub-exposures. To address this, we recalculated the errors for each sub-exposure using the {\tt GROSS} counts at each wavelength element, shown in Equation~\ref{eq:error}, which is the approximate $1-\sigma$ confidence limit of a Poisson distribution. The ratio of {\tt GROSS} to {\tt FLUX} gave us a count-to-flux-units conversion factor for each wavelength element.

\begin{equation} \label{eq:error}
    \sigma \approx 1 + \sqrt{N + 0.75}
\end{equation}

\subsection{Breathing correction for new observations} \label{sec:breathe}

\begin{figure*}[ht!]
\epsscale{1.15}
\plotone{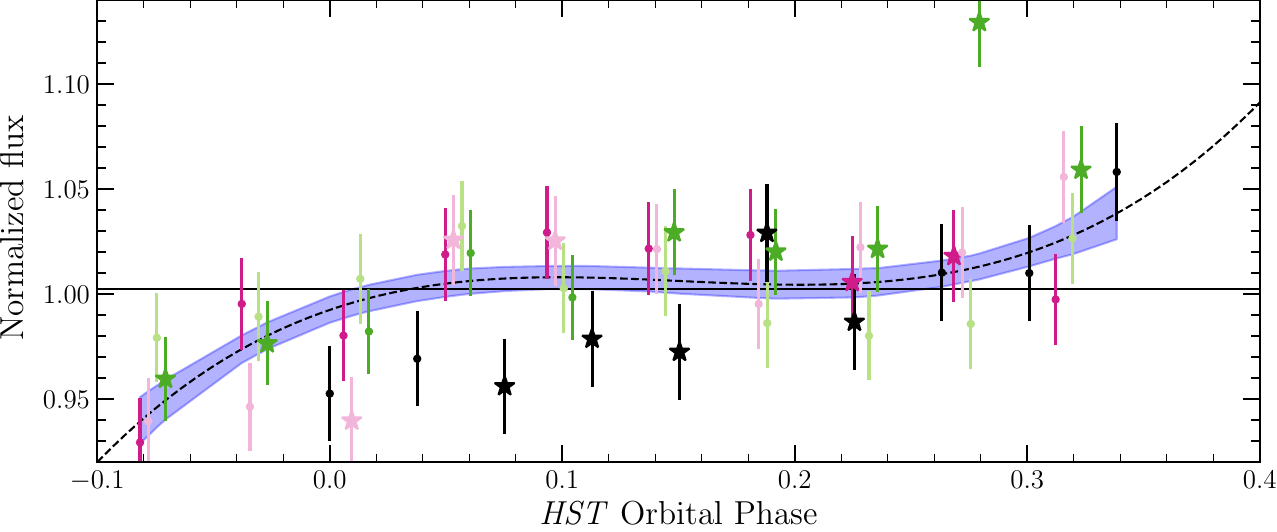}
\caption{Our sub-exposure \lya\ light curves folded on {\it HST}'s orbital period of $95.42$ minutes. Each color corresponds to one full exposure (one orbit) and each point corresponds to a sub-exposure normalized by the average flux of each orbit. The dashed line depicts the best-fit polynomial as determined by our method outlined in Section~\ref{sec:breathe} compared to a flat line. The blue shaded region represents the $1-\sigma$ uncertainties in the fit evaluated at each sub-exposure. The star data points represent the flaring sub-exposures that were not considered when fitting for breathing. \label{fig:breathing}}
\end{figure*}

Thermal fluctuations of {\it HST} can lead to systematic flux changes over the course of an orbit (``breathing"). We searched for in-orbit trends by folding each exposure on the length of an {\it HST} orbit ($95$ minutes). We tested several polynomial fits to the folded light curves of the fully-integrated \lya\ line ($1214 - 1215.4$ \AA\ and $1215.7 - 1217$ \AA, which avoids core geocoronal emission and is the highest signal-to-noise region of the observed spectrum), ignoring exposures that coincided with flares (see Section~\ref{sec:flares}). The best-fit trend was determined by the polynomial fit that gave the smallest Bayesian Information Criterion (BIC). Figure~\ref{fig:breathing} depicts the best-fit breathing trend in our data; the phase is arbitrary. We removed this instrumental effect by dividing the folded light curve by the best-fit polynomial. We added the $1-\sigma$ uncertainty in the breathing trend in quadrature to the statistical errors. Because breathing should affect all wavelengths similarly, the same polynomial was divided out of all of the FUV emission line light curves.

\subsection{Archival FUV observations of AU Mic}

We compare our transit data to archival AU Mic observations \citep{2000ApJ...532..497P}. The archival data consists of four consecutive exposures from September 1998 (HST-GO-7556; PI: Linsky) taken with the same observing set-up as our data. We followed the same reduction and extraction procedure as described in Section~\ref{sec:newdata}. The observation times of these data do not correspond to a transit of either AU Mic b or c when propagating their TESS-derived periods and errors back in time. The nearest AU Mic b and c transits to the September 1998 data are $58 \pm 1.3$ hours after and $15 \pm 0.74$ hours before, respectively, with propagated uncertainties from \cite{2023AJ....166..232W} without accounting for TTVs.

We also compare our AU Mic c transit data to the archival AU Mic b transit data from Visits 1 (2 July 2020) and 2 (19 October 2021) in \cite{2023AJ....166...77R}, which correspond to the non-detection and detection of AU Mic b's exosphere, respectively. Table~\ref{tab:obs} is an observation log for all successful and failed observations discussed in this work. Figure~\ref{fig:lyaspec} shows the \lya\ spectra for all four AU Mic epochs referenced in this work.

\section{AU Mic's FUV Light Curves} \label{sec:lcurve}

\begin{figure*}
\epsscale{1.15}
\plotone{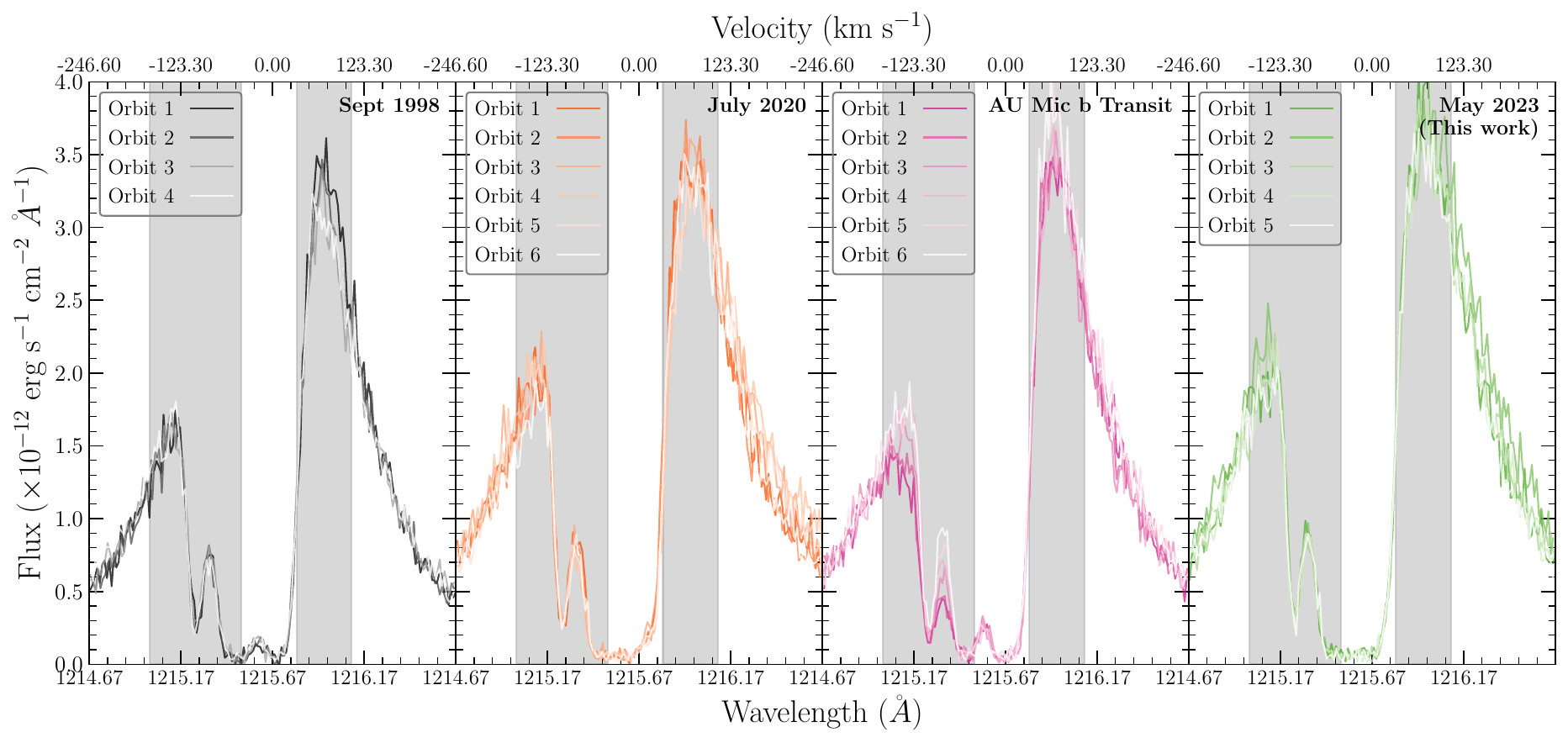}
\caption{AU Mic's \lya\ emission line observed on September 1998 (left; no transit; orbits are 56-63 hours after AU Mic b mid-transit), July 2020 (second-from-left; undetected AU Mic b transit; orbits are 6 hours prior to 2 hours after transit midtime with one final orbit 17 hours after transit), October 2021 (third-from-left; detected AU Mic b transit; orbits are 4 hours prior to 4 hours after mid-transit) and May 2023 (right; during AU Mic c transit; orbits are 3 hours prior to 4 hours after AU Mic c mid-transit). The gray shaded regions on the left and right of each panel correspond to the integrated blue- and red-wing regions, respectively. The region in between the wings is where geocoronal contamination and ISM absorption occurs. The legends specify each orbit in the four visits.} \label{fig:lyaspec}
\end{figure*}

\begin{figure*}
\epsscale{1.15}
\plotone{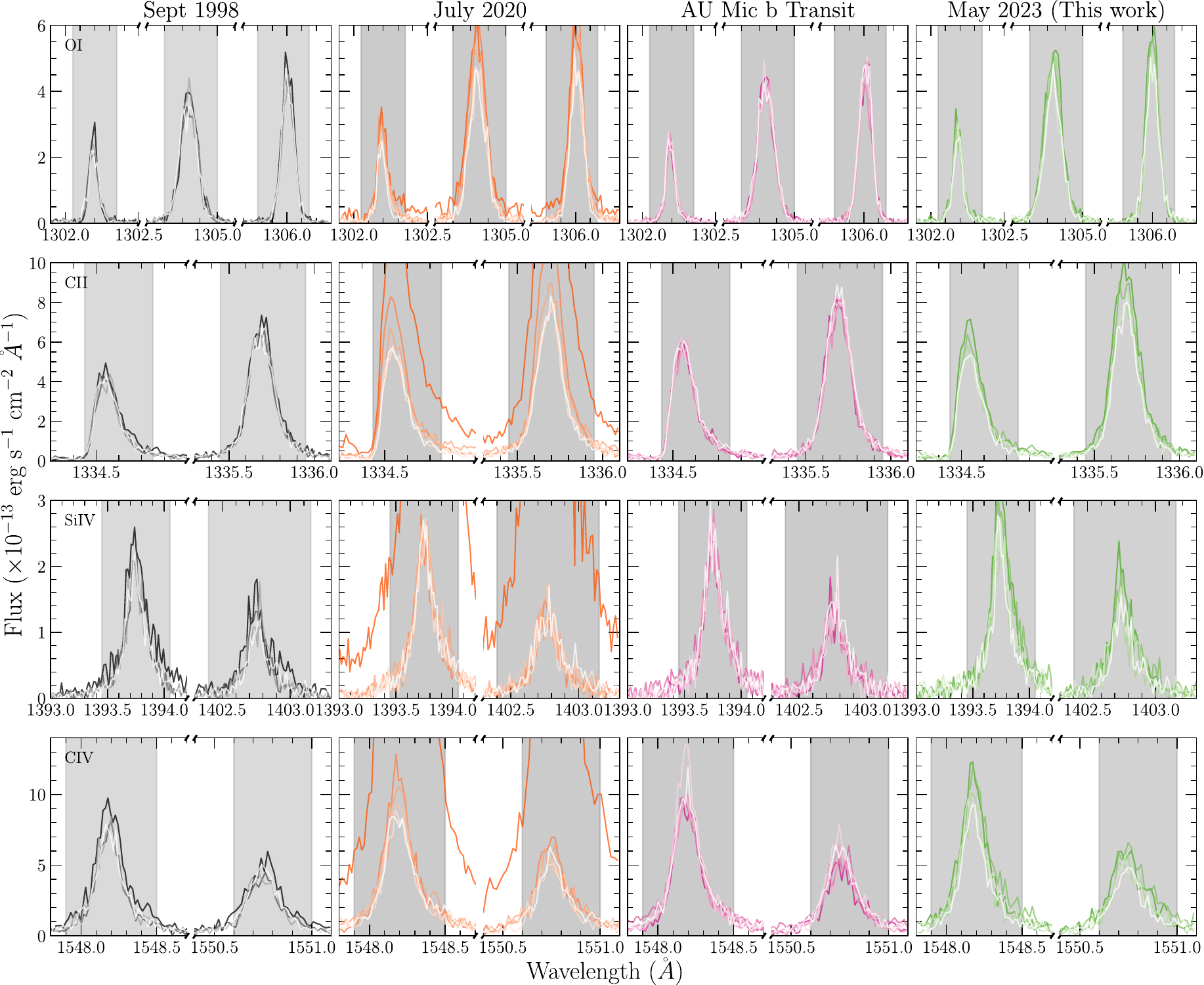}
\caption{AU Mic spectra observed by {\it HST}/STIS on September 1998 (left; no transit), July 2020 (second-from-left; undetected AU Mic b transit that also displayed a large flare), and October 2021 (third-from-left; detected AU Mic b transit) and May 2023 (right; during AU Mic c transit). \ion{O}{1} and \ion{C}{2} are shown to search for planetary absorption. \ion{Si}{4} and \ion{C}{4} are also shown as the FUV emission lines {\it most sensitive} to stellar activity in this work. The gray regions show the wavelength ranges integrated over to obtain the light curves in Figure~\ref{fig:lcurve}. Each line color corresponds to same orbit from Figure~\ref{fig:lyaspec}. All other FUV emission lines from Table~\ref{tab:lines} can be found in Figure~\ref{fig:other_spec}. \label{fig:spec}}
\end{figure*}

The emission lines analyzed in this work are listed in Table~\ref{tab:lines}. We show the subset of lines we discuss in Figures \ref{fig:lyaspec} and \ref{fig:spec}, and include the remainder in \ref{fig:other_spec}. We created light curves for each of the emission lines. The line fluxes were integrated over the wavelength ranges indicated by the gray regions in Figures \ref{fig:lyaspec}, \ref{fig:spec}, and \ref{fig:other_spec}. Their respective uncertainties were calculated by summing the errors at each wavelength element in quadrature.

\begin{deluxetable*}{cllll}[t!]
\tablecaption{AU Mic FUV emission lines identified in {\it HST}/STIS spectra and used in this work. Line location(s), integration region, and formation temperature are listed \citep{1997AAS..125..149D,2021ApJ...909...38D}. We also indicate whether the line was used to search for stellar (S) or planetary (P) trends. \label{tab:lines}}
\tablecolumns{5}
\tablewidth{0pt}
\tablehead{
\colhead{Species} &
\colhead{$\lambda$ (\AA)} &
\colhead{$\Delta\lambda$ (\AA)} &
\colhead{$\log_{10}(T/\text{K})$} &
\colhead{S/P}
}
\startdata
\ion{C}{3} & $1175.7$ & $1174.5 - 1177.0$ & $4.8$ & S \\
\ion{H}{1} (\lya) & $1215.67$ & $1214.0 - 1217.0$ & $4.5$ & P \\
\ion{N}{5} & $1238.82$, $1242.80$ & $1238.6 - 1239.0$, $1242.6 - 1243.0$ & $5.2$ & S \\
\ion{O}{1} & $1302.2$, $1304.84$, $1306.01$ & $1302.05 - 1302.35$, $1304.7 - 1305.0$, $1305.8 - 1306.15$ & $3.8$ & P \\
\ion{C}{2} & $1334.53$, $1335.71$ & $1334.45 - 1334.75$, $1335.45 - 1335.95$ & $4.5$ & P \\
\ion{Si}{4} & $1393.76$, $1402.77$ & $1393.45 - 1394.05$, $1402.4 - 1403.15$ & $4.9$ & S \\
\ion{C}{4} & $1548.20$, $1550.774$ & $1547.9 - 1548.5$, $1550.6 - 1551.0$ & $4.8$ & S \\
\ion{He}{2} & $1640.40$ & $1640.2 - 1640.65$ & $4.9$ & P \\
\ion{C}{1} & \begin{tabular}{l}$1656.27$, $1656.93$, $1657.01$, $1657.38$, \\$1657.91$, $1658.12$\end{tabular} & $1656.0 - 1658.25$ & $3.8$ & P \\
\enddata
\end{deluxetable*}
We calculated our ephemeris from the AU Mic c transit ephemeris and propagated period from \cite{2023AJ....166..232W}. As shown in Figure~\ref{fig:ephem}, this is in good agreement with the most-favored TTV model from \cite{2023AJ....166..232W} and we expect that our data covers the AU Mic c transit in spite of the TTV uncertainty.

For photoevaporation-driven planetary outflows, the escaping material is launched at $10$s of km s$^{-1}$ \citep[e.g.,][]{2009ApJ...693...23M,2018ApJ...855L..11O,2022A&A...659A..62D}. If AU Mic c has a substantial outflow of neutral hydrogen that is accelerated to $> \pm100$ km s$^{-1}$ in AU Mic's rest-frame (via external mechanisms such as radiation pressure, stellar wind ram pressure, and charge exchange with the stellar wind), it will absorb parts of the \lya\ line wings while in transit. While there is uncertainty over the acceleration mechanism(s), such high velocity material is observed for other planets \citep[e.g.,][]{2017A&A...605L...7L,2018A&A...620A.147B} and predicted by models \citep[e.g.,][]{2013A&A...557A.124B,2019ApJ...873...89M,2019ApJ...885...67K}. Additionally, an optically thick neutral hydrogen exosphere could absorb the \lya\ wings without being accelerated \citep[e.g.,][]{2019ApJ...873...89M}. The \lya\ blue- and red-wings examined in this work cover $-165 < v < -42$ and $32 < v < 106$ km s$^{-1}$, respectively, and are indicated by the gray regions in Figure~\ref{fig:lyaspec}. The former correspond to the wavelengths over which the AU Mic b transit was detected. We chose these integration windows for our \lya\ light curves to maximize signal while minimizing stellar contamination from the outer wings. The low velocity material ($10$s of km s$^{-1}$) is not observable due to absorption of the line core by the interstellar medium (ISM). We note that any geocoronal contamination --- excess emission in between the gray regions in Figure~\ref{fig:lyaspec} --- is confined to the ISM absorption region and is not a limiting factor in our analysis.

\cite{2023AJ....166...77R} published observations of AU Mic b's exosphere, which can be seen absorbing up to $\sim 30$\% of the \lya\ blue-wing in the third panel from the left in Figure~\ref{fig:lyaspec}. The absorbing material is observed across multiple orbits, and is observed only at specific blue-shifted velocities (gray region). We do not see similar absorption for AU Mic c, as shown in Figure~\ref{fig:lyaspec}'s right-most panel and the \lya\ blue- and red-wing light curves presented in Figure~\ref{fig:lcurve} (to be discussed in Section~\ref{sec:pl_search}).

We also looked for trends in other FUV emission lines (see Figures~\ref{fig:spec} and~\ref{fig:other_spec}). Metal species in lower ionization states, like \ion{O}{1} and \ion{C}{2}, could be tracers for atmospheric escape as they are dragged off the planet by the lighter species' outflow. \cite{2024RNAAS...8...86F} did not detect an exosphere in their FUV metal transmission observations of young planets AU Mic b and V1298 Tau c. Several other searches for metal absorption have been conducted with $\pi$~Men c \citep{2021ApJ...907L..36G}, HAT-P-11 b \citep{2022NatAs...6..141B} and the hot Jupiters WASP-121 b and HD 189733 b and more \citep{2019AJ....158...91S,2023AJ....166...89D}. $\pi$~Men c and HAT-P-11 b remain the only planets smaller than Jupiter with detected metal exospheres.

We compared the behavior of the \ion{O}{1} and \ion{C}{2} lines to the behavior of the lines most sensitive to stellar activity in our AU Mic dataset -- \ion{Si}{4} and \ion{C}{4} -- in order to disentangle the flux changes caused by the star from those potentially caused by intervening planetary material. All emission lines, excepting the wings of \lya, show a decrease in flux from the first orbit to the last around the AU Mic c transit, which can be seen in the spectra in Figure~\ref{fig:spec} (additional lines can be found in \ref{fig:other_spec}). In the following sections, we seek to distinguish whether this behavior is planetary or stellar in nature.

\begin{figure*}
\epsscale{1.15}
\plotone{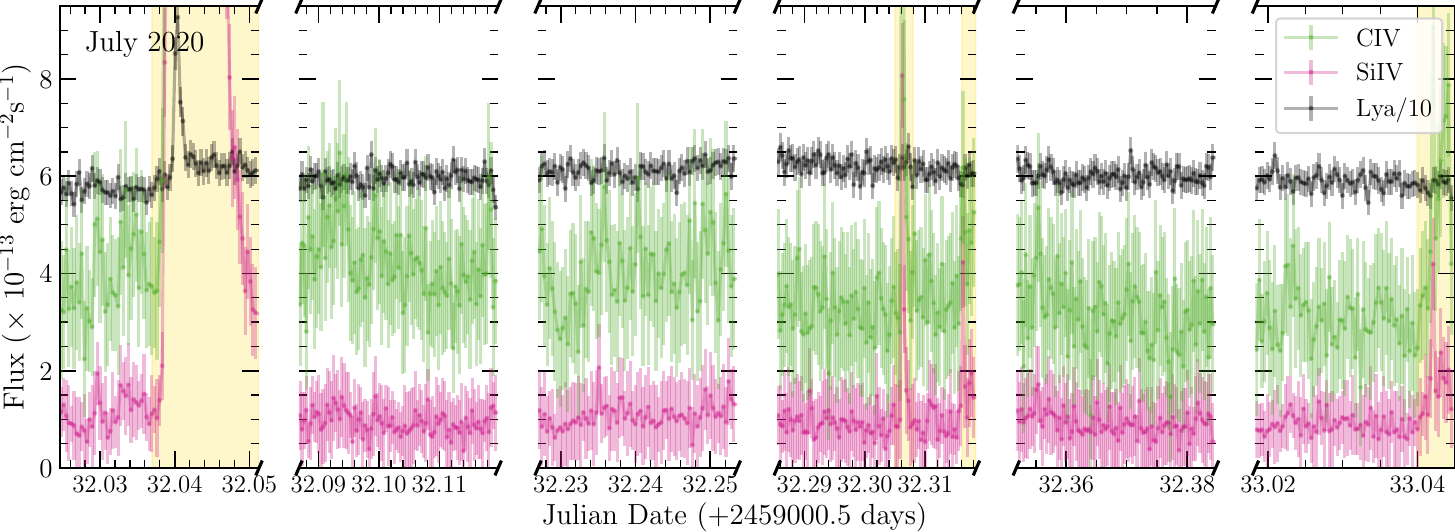}
\plotone{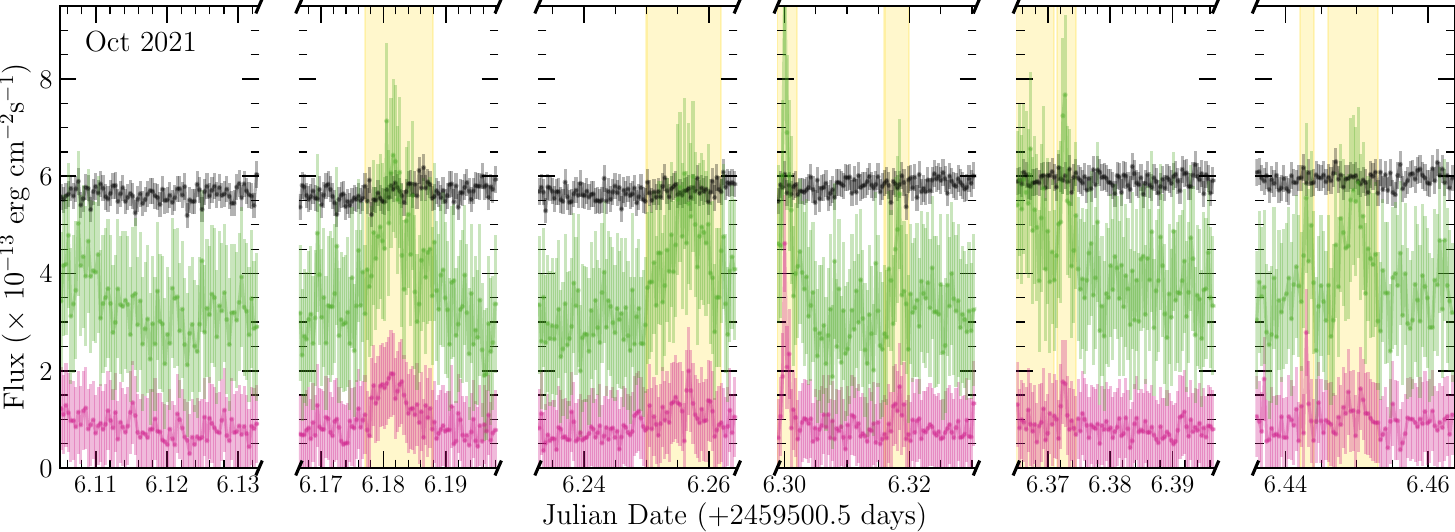}
\plotone{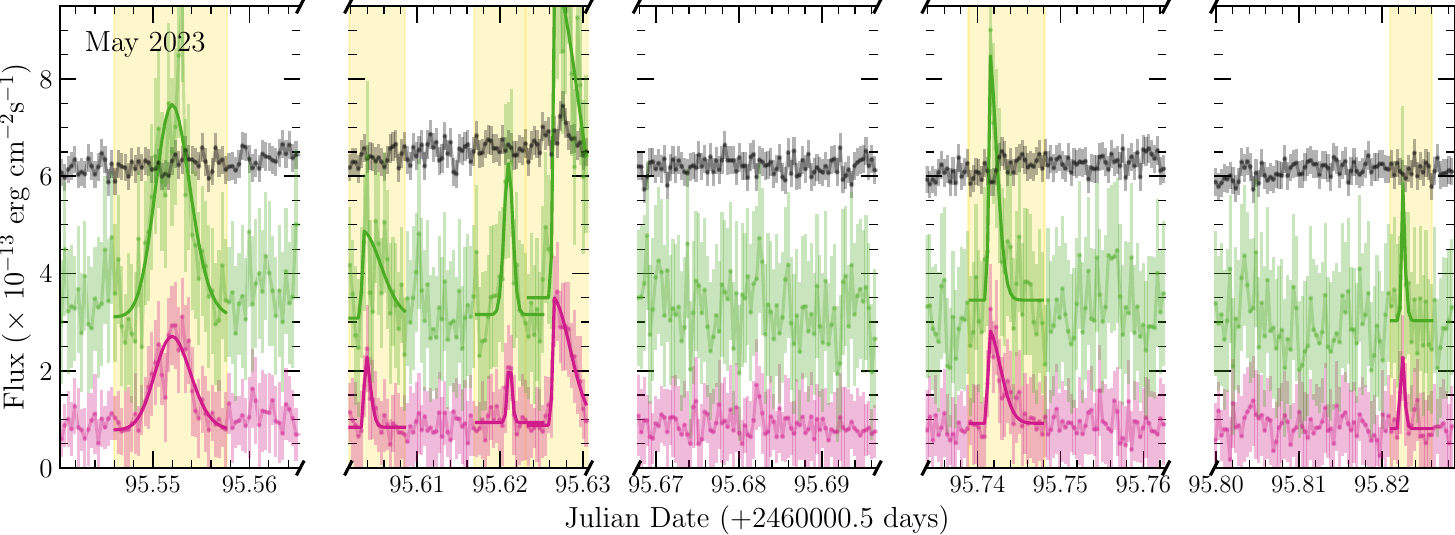}
\caption{AU Mic \ion{C}{4}, \ion{Si}{4}, and \lya\ (divided by 10 plus 3) emission over time during the July 2020 (top; AU Mic b transit), October 2021 (middle; AU Mic b transit), and May 2023 (bottom; AU Mic c transit) visits. Flares are highlighted in gold. The best-fit flare models for the May 2023 flares are shown in bold. Figure~\ref{fig:cont_flares} contains similar FUV continuum light curves.}
\label{fig:flares}
\end{figure*}

\subsection{Search for flares} \label{sec:flares}

We further broke up the data into $30$s sub-exposures and searched for flares using light curves for all emission lines listed in Table~\ref{tab:lines}. Flares were identified by eye in parts of the light curve where at least three consecutive sub-exposures are above the noise level of the surrounding quiescent flux. \ion{Si}{4} and \ion{C}{4} presented as the most sensitive and so were used for flare identification. The \ion{Si}{4} and \ion{C}{4} light curves are compared to AU Mic's \lya\ emission (divided by 10) in Figure~\ref{fig:flares}. We include the FUV continuum light curves in Section~\ref{ap:cont} for reference, but chose to emphasize the emission line light curves as they are more sensitive to flares in this dataset. The two previous visits to AU Mic --- July 2020 and October 2021, both coinciding with an AU Mic b transit --- are included for comparison. We do not consider the September 1998 observations in this section, although some flaring behavior can be seen in the \ion{Si}{4} and \ion{C}{4} light curves at the bottom of Figure~\ref{fig:lcurve}, because the star varies more significantly across decades than across years.

One exceptionally high-energy flare was observed in the July 2020 visit and caused an increase in AU Mic's \lya\ emission, which is observable in the time series data in Figure~\ref{fig:flares}. \lya\ decayed much quicker than \ion{Si}{4} and \ion{C}{4} and the \lya\ flux increase itself is not expected to have prevented the detection of a neutral hydrogen exosphere.  However, \cite{2023AJ....166...77R} proposed that the concurrent ionizing flux from the flare may have ionized the exosphere such that there was little to no neutral hydrogen present at that time. Over the rest of the July 2020 and October 2021 visits, we find 11 lower-energy flares (highlighted in gold in Figure~\ref{fig:flares}). None correspond to increased \lya\ emission.

For the May 2023 data presented in this work, we see 6 distinct low-energy flares in \ion{Si}{4} and \ion{C}{4}. We modeled and fit each May 2023 flare using {\tt cos\_flares} from \cite{2022AJ....164..110F} modified to take STIS data. All flares were fit using a skewed Gaussian model (Equation 1 from \citealt{2022AJ....164..110F}). The best-fit equivalent durations (Equation 5 from \citealt{2022AJ....164..110F}) for the May 2023 \ion{Si}{4} and \ion{C}{4} light curves are listed in Table~\ref{tab:flares}. None of the flares in either line lasted longer than $14.6$ minutes. We do not know the impact of the different instrument throughputs on flare detection and modeling and therefore do not directly compare our flares with those found in \cite{2022AJ....164..110F} with {\it HST}/COS.

Across the three {\it HST} visits from 2020 -- 2023, 18 flares in total were observed over $12.1$ hours, which equates to 1 flare every $40$ minutes. This is similar to the $\sim 2$ flares per hour flare rate for AU Mic estimated by \cite{2022AJ....164..110F} using {\it HST}/COS. For comparison, \cite{2022AJ....163..147G} detected 5.5 flares/day in {\it TESS} 20-second white-light curves of AU Mic, while \cite{2023ApJ...951...33T} detected 10 flares/day with brightness $>10^{31}$ ergs in U-band photometry (see their Figure 10).

Lower-energy flares that do not significantly impact AU Mic's \lya\ emission are more common (only 2/18 flares are visible in \lya). The morphologies of the same flare in different FUV emissions lines are consistent, which was not the case for flares observed in \cite{2022AJ....164..110F} and \cite{2014ApJS..211....9L}. However, there are morphological differences when comparing different flares for the same emission line. For example, the first flare in the May 2023 visit exhibits a steady rise and decay that differs from the canonical quick rise and steady decay seen in the second, fourth and fifth flares of the same visit. This could result from a phenomenon similar to the ``peak-bump'' morphology discussed in precise, broadband optical photometry from \cite{2022ApJ...926..204H}; in this work the ``bump'' was occasionally observed without the ``peak.'' Alternatively, and we consider more likely, is that the observed flare morphology derives from an underlying complex flare composed of multiple components. This could produce the more symmetric shapes we see when sampled at the cadence and signal-to-noise of our observations.

Only one flare from the July 2020 visit, during the first orbit, and one in the May 2023 visit, during the second orbit, corresponded to a \lya\ flare. We modeled and fit the July 2020 and May 2023 \lya\ flares with the same methods used for the \ion{Si}{4} and \ion{C}{4} flares, using a single skewed Gaussian profile. We measure the July 2020 and May 2023 flare energies to be $6.29 \times 10^{30}$ ergs and $3.93 \times 10^{30}$ ergs, respectively. Their equivalent durations are $192$s and $108$s.

We used the best-fit \lya\ flare models to measure how long each flare would mask a $10$\% hypothetical transit. The July 2020 \lya\ flare would mask a $10$\% hypothetical transit for $840$s, and the May 2023 \lya\ flare did the same for about $510$s. Both flares return to $<10$\% over quiescent flux in less than an hour, which is similar to the short durations found in \cite{2014ApJS..211....9L}. While this increase in \lya\ flux does mask the presence of neutral hydrogen in transit at the $10$\% level for this short period of time, it is still possible that the canonical picture of atmospheric escape off of close-in planets --- i.e., a large trailing tail that absorbs \lya\ for hours after transit --- is detectable. We note our July 2020 flare measurements are different from those reported in \cite{2023AJ....166...77R} because we chose to fit a smaller window of time; we determined that fitting the larger window of time spanning two orbits inflated the length of the flare.

\begin{deluxetable}{ccccc}
\tablecaption{Best-fit equivalent durations (s) for all \lya, \ion{Si}{4} and \ion{C}{4} flares detected in the May 2023 visit to AU Mic. The \lya\ effective duration (s) is also included; this is the time to return to $10$\% of baseline flux, corresponding to the length of time for which a $10$\% hypothetical transit would be masked by the transit. \label{tab:flares}}
\tablecolumns{5}
\tablewidth{0pt}
\tablehead{
\colhead{Flare} &
\colhead{\ion{Si}{4}} &
\colhead{\ion{C}{4}} &
\colhead{\lya} &
\colhead{\lya\ (effective)}
}
\startdata
1 & $874$ & $458$ & --- & --- \\
2 & $146$ & $95.1$ & --- & --- \\
3 & $186$ & $70.2$ & --- & --- \\
4 & $740$ & $583$ & $108$ & $510$ \\
5 & $470$ & $209$ & --- & --- \\
6 & $93.7$ & $6.31$ & --- & --- \\
\enddata
\end{deluxetable}

\subsection{Impact of flares on the search for planetary absorption}

We hypothesize that the high level of activity in the first two orbits of May 2023 ---  seen in the \lya\ flare in conjunction with the other emission line flares around the same time --- are the cause for the noticeable $\sim10\%$ decrease in the average \lya\ flux between the second and third orbits of the May 2023 visit seen in Figure~\ref{fig:flares}. Whereas a planetary cause is expected to be wavelength-dependent due to the dynamics of the exospheric material, this decrease is experienced by the {\it entire} \lya\ line (similar behavior at all wavelengths included in both the blue- and red-wing light curves in Figure~\ref{fig:lcurve}). The first orbit is also similar in flux to the orbits after the flare, which points towards an increase in \lya\ flux associated with the flares in the second orbit followed by a decline back to quiescence by the third orbit. This is consistent with the short duration we measured for the \lya\ flare.

While the \lya\ flux increase in the second orbit proves inconvenient to conclusively search for planetary material in transit, a large transiting neutral hydrogen exosphere like that of AU Mic b ($>10\%$ transit depth at blue-shifted velocities) should still dominate the signal in our light curves {\it if present}. However, if one or all of the May 2023 flares corresponded to significantly increased ionizing radiation ($<912$ \AA), then any neutral hydrogen escaping AU Mic c may be ionized and rendered unobservable for some period of time over the course of the transit. We do not yet know the rapidity, intensity and duration of a single flare's impact on the observability of hydrogen exospheres, although increasing amounts of literature point towards the importance of constraining this impact by reconciling simulations and observations \citep[e.g.,][]{2014ApJS..211....9L,2022AJ....164..110F,2023MNRAS.518.4357O,2023AJ....166...77R}.

The full exposures for our May 2023 light curves in Figure~\ref{fig:lcurve} only contain co-added non-flaring sub-exposures. For illustrative purposes, we leave the flaring sub-exposures in all light curves. The flaring regions were not considered during the searches for {\it HST} breathing or planetary material (Sections~\ref{sec:breathe} and~\ref{sec:pl_search}).

\subsection{Search for planetary absorption at FUV emission lines} \label{sec:pl_search}

With careful consideration of the stellar activity discussed in the previous sections, we look for signatures of planetary absorption at \lya\ and two metal species, \ion{O}{1} and \ion{C}{2}.

\begin{figure*}
\epsscale{1.1}
\plotone{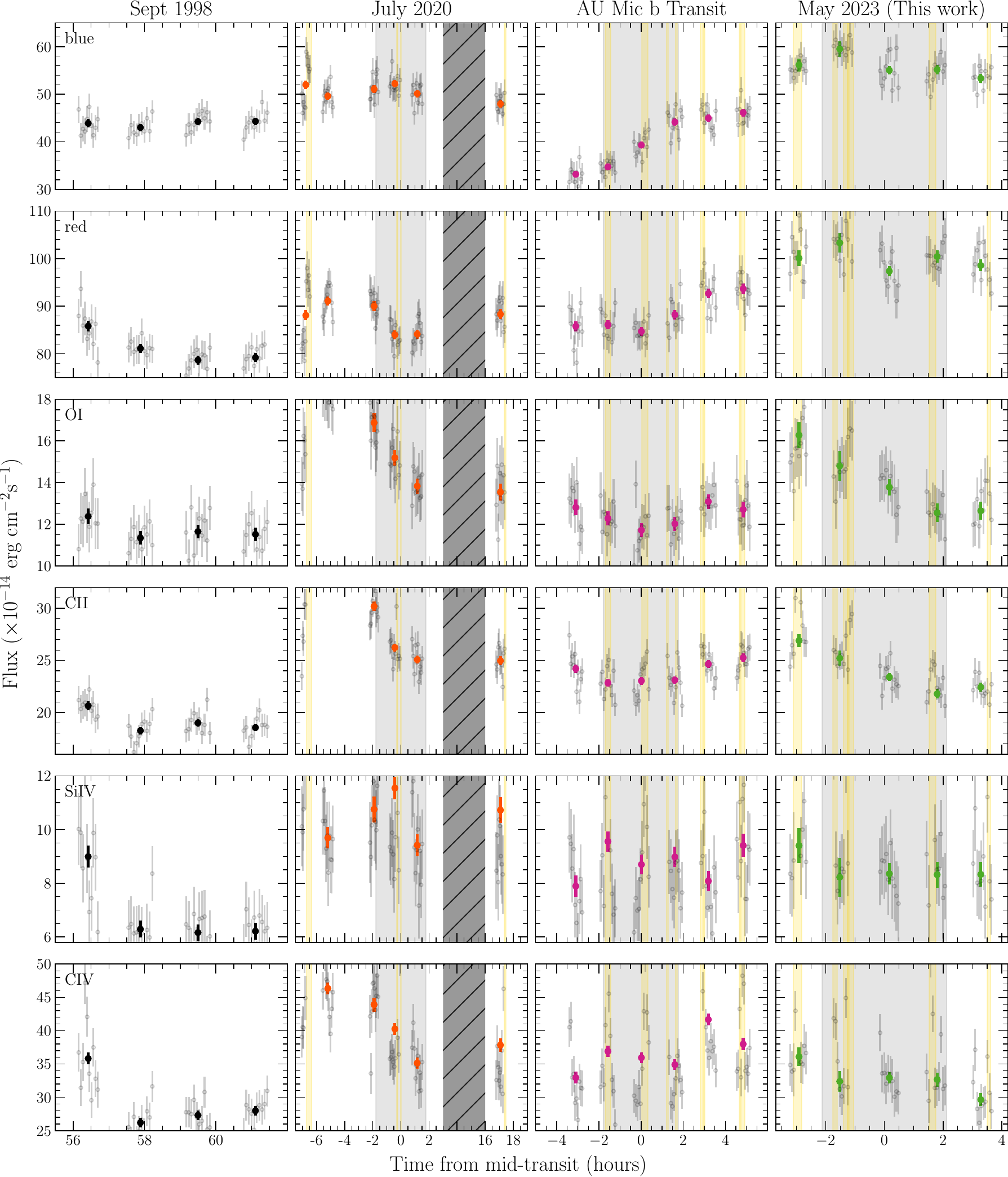}
\caption{AU Mic light curves observed by {\it HST}/STIS on September 1998 (left; no transit but plotted in reference to AU Mic b ephemeris), July 2020 (second-from-left; undetected AU Mic b transit), October 2021 (third-from-left; detected AU Mic b transit) and May 2023 (right; during AU Mic c transit). The \lya\ blue- and red-wings, \ion{O}{1} and \ion{C}{2} are shown to search for planetary absorption. \ion{Si}{4} and \ion{C}{4} are also shown as the FUV emission lines {\it most sensitive} to stellar activity in this work. The gray regions indicate the white-light transit durations of AU Mic b and c. Flares are highlighted in gold. All other FUV emission lines from Table~\ref{tab:lines} can be found in Figure~\ref{fig:other_lcurve}. \label{fig:lcurve}}
\end{figure*}

As indicated by the spectra themselves, the \lya\ blue- and red-wing light curves (top two rows of Figure~\ref{fig:lcurve}) show no indication of absorption from planetary neutral hydrogen during AU Mic c's transit. The level of variability between the archival light curves with no planetary absorption and the May 2023 light curves analyzed here are very similar. There is no drastic wavelength-dependent flux decrease in the May 2023 light curves, as was seen during (and only during) the AU Mic b transit in the \lya\ blue-wing. As discussed in Section~\ref{sec:flares}, which sifted through higher temporal sampling of AU Mic observations to look for short term variability, flares that occurred during our May 2023 observations are not significant enough to impact the orbit-integrated \lya\ spectra (right-most column of Figures \ref{fig:lyaspec}).

We visually compared the light curves for metal species expected in planetary exospheres, \ion{O}{1} and \ion{C}{2}, to those most sensitive to stellar flares, \ion{Si}{4} and \ion{C}{4}. We find that \ion{O}{1} and \ion{C}{2} exhibit similar light curve behavior to each other. For example, in the July 2020 data, \ion{O}{1} and \ion{C}{2} fluxes drastically increase during the very large flare within the first orbit, and that flux increase persists longer than in other emission lines. Further, the \ion{O}{1} and \ion{C}{2} flux changes in the May 2023 data occur in both the negative and positive velocities of each line's profile. Absorption (at the $>10$\% level) due to intervening escaping material is expected to produce asymmetric flux changes, as was seen for the detection of \ion{C}{2} in HAT-P-11 b's exosphere \citep{2022NatAs...6..141B}. Thus, we attribute the \ion{O}{1} and \ion{C}{2} variability to stellar activity. Smaller absorption signals, which we cannot probe in this work, would be indicative of material at lower altitudes. This material would behave more like a Parker wind and show symmetric depreciation in wavelength.

Because the \ion{O}{1} and \ion{C}{2} light curves in Figure~\ref{fig:lcurve} do exhibit some ``by eye" transit-like behavior in our May 2023 data, we quantitatively explore this variability. We tested whether the in-transit fluxes differed significantly from the out-of-transit fluxes. We defined in-transit as all exposures taken during the May 2023 visit (AU Mic c transit), and out-of-transit as the last three exposures from the July 2020 visit (away from the flare identified in \citealt{2023AJ....166...77R}) and the last two exposures from the October 2021 visit (outside of the AU Mic b transit). The in- and out-of-transit sub-exposure flux distributions were resampled and put through a two-sample Kolmogorov-Smirnov test. This was repeated for each emission line light curve pictured in Figures~\ref{fig:lcurve} and~\ref{fig:other_lcurve}. The flux distributions for both in- and out-of-transit were not significantly different (all p-values were $\sim0.15 - 0.65$), meaning that the May 2023 data did not significantly differ from the archival data.

\subsection{Long-term Lyman-alpha variability}

Having established that AU Mic c is likely not impacting the May 2023 \lya\ spectra, we can look at long-term changes in AU Mic's \lya\ emission. We reconstructed AU Mic's intrinsic (not contaminated by the ISM) \lya\ emission across all four visits using {\tt lyapy} \citep{2016ApJ...824..101Y,2022zndo...6949067Y}. We modeled the observed \lya\ profile as a double-Gaussian stellar emission component combined with an ISM Voigt absorption component. The double-Gaussian component is a superposition of a narrow and a broad Gaussian profile described by their velocity centroids (v$_{\text{n}}$, v$_{\text{b}}$), amplitudes (A$_{\text{n}}$, A$_{\text{b}}$), and full-width half-maxmima (FW$_{\text{n}}$, FW$_{\text{b}}$). The interstellar medium's Voigt profile is characterized by a single cloud column density (log$_{10}$N$_{\text{H I}}$), Doppler broadening (b), velocity (v$_{\text{H I}}$), and deuterium-to-hydrogen ratio. We are not looking to accurately measure ISM parameters and so do not consider multiple cloud components. The best-fit components and resulting intrinsic profile is found with the MCMC algorithm {\tt emcee} \citep{2013PASP..125..306F}.

We chose the last orbit in each visit (darkest profiles in Figure~\ref{fig:lyaspec}) to reconstruct AU Mic's intrinsic \lya\ emission as they are the farthest away from any potential planet transit. We considered all stellar and ISM parameters as free with the exception of the deuterium-to-hydrogen ratio, which was fixed to the literature-accepted value of $1.5\times 10^{-5}$ \citep{2006ApJ...647.1106L}. We ran the algorithm the same for each AU Mic visit -- uniform priors for all parameters, ran for 20,000 steps, 30 walkers, and 4,000 burn-in. The best-fit parameters for all visits are listed in Table~\ref{tab:lyapy}.

\begin{deluxetable*}{lrrrr}
\tablecaption{Best-fit parameters with $1-\sigma$ uncertainties for the four AU Mic visits, characterizing the intrinsic \lya\ line and the ISM absorption. \label{tab:lyapy}}
\tablecolumns{5}
\tablewidth{0pt}
\tablehead{
\colhead{Parameter (Units)} &
\colhead{Sept 1998} &
\colhead{July 2020} &
\colhead{Oct 2021} &
\colhead{May 2023}
}
\startdata
v$_{\text{n}}$ (km s$^{-1}$) & $-9.38^{+0.92}_{-0.89}$ & $-7.98^{+0.90}_{-0.81}$ & $-7.29^{+0.92}_{-0.89}$ & $-7.27^{+1.00}_{-0.95}$ \\
log$_{10}$A$_{\text{n}}$ (erg s$^{-1}$ cm$^{-2}$ \AA$^{-1}$) & $-10.94^{+0.03}_{-0.03}$ & $-10.94^{+0.02}_{-0.03}$ & $-10.87^{+0.03}_{-0.03}$ & $-10.98^{+0.03}_{-0.02}$ \\
FW$_{\text{n}}$ (km s$^{-1}$) & $144.57^{+3.19}_{-3.08}$ & $153.59^{+2.85}_{-2.93}$ & $140.54^{+2.82}_{-2.74}$ & $151.63^{+3.06}_{-2.97}$\\
v$_{\text{b}}$ (km s$^{-1}$) & $-15.77^{+1.20}_{-1.20}$ & $2.43^{+1.76}_{-1.63}$ & $-7.37^{+1.18}_{-1.18}$ & $-9.74^{+1.22}_{-1.18}$ \\
log$_{10}$A$_{\text{b}}$ (erg s$^{-1}$ cm$^{-2}$ \AA$^{-1}$) & $-11.69^{+0.02}_{-0.02}$ & $-11.73^{+0.02}_{-0.02}$ & $-11.64^{+0.02}_{-0.02}$ & $-11.62^{+0.02}_{-0.02}$ \\
FW$_{\text{b}}$ (km s$^{-1}$ ) & $377.79^{+6.55}_{-5.97}$ & $428.27^{+8.34}_{-8.49}$ & $386.12^{+6.03}_{-5.87}$ & $391.80^{+6.09}_{-5.81}$ \\
ISM log$_{10}$N$_{\text{H I}}$ & $18.40^{+0.01}_{-0.01}$ & $18.38^{+0.01}_{-0.01}$ & $18.38^{+0.01}_{-0.01}$ & $18.38^{+0.01}_{-0.01}$ \\
b (km s$^{-1}$) & $12.42^{+0.33}_{-0.33}$ & $12.09^{+0.32}_{-0.38}$ & $12.34^{+0.27}_{-0.29}$ & $11.16^{+0.33}_{-0.38}$ \\
v$_{\text{H I}}$ (km s$^{-1}$) & $-23.42^{+0.25}_{-0.25}$ & $-22.75^{+0.25}_{-0.27}$ & $-22.98^{+0.23}_{-0.23}$ & $-24.42^{+0.24}_{-0.23}$ \\
\enddata
\end{deluxetable*}

\begin{figure}
\epsscale{1.15}
\plotone{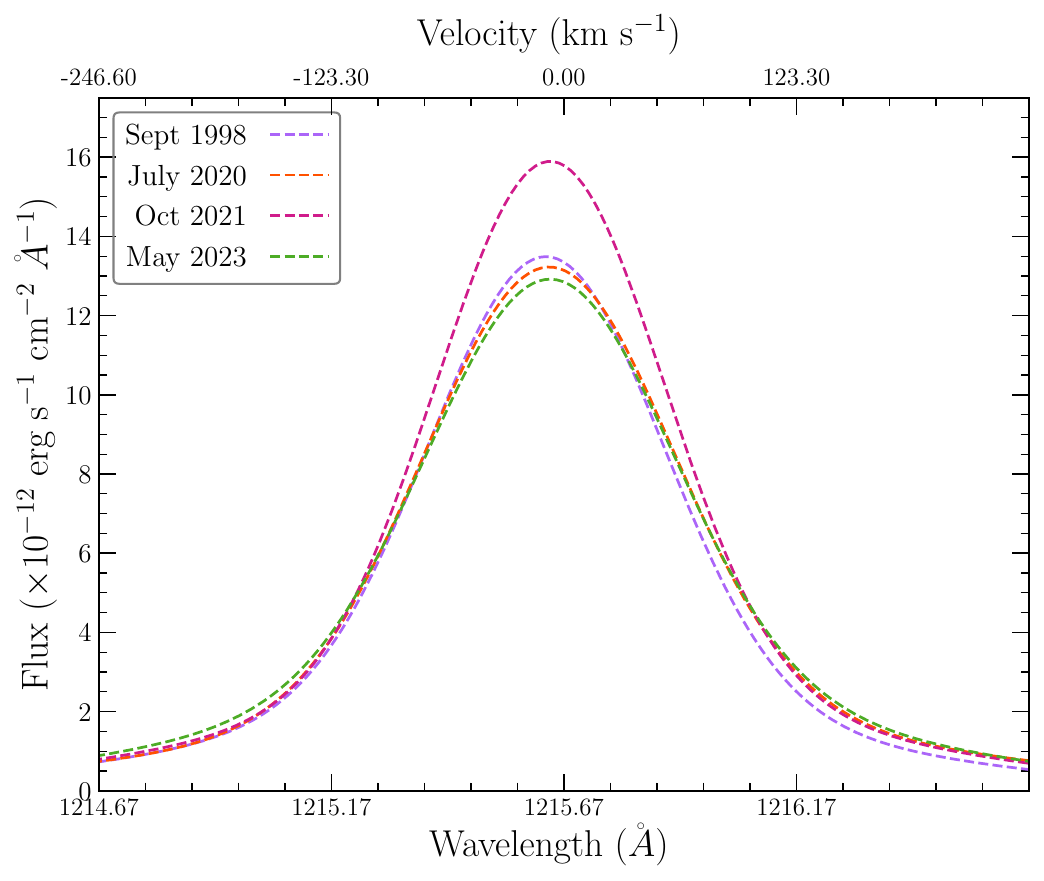}
\caption{AU Mic's reconstructed intrinsic \lya\ emission line for the last orbit of the September 1998 (purple), July 2020 (orange), October 2021 (pink) and May 2023 (green) visits. \label{fig:lya_profs}}
\end{figure}

Although the intrinsic profiles (Figure~\ref{fig:lya_profs}) do differ from each other, the integrated \lya\ emission (and ISM column density) for each visit are in remarkable agreement. The resulting \lya\ fluxes and their $1-\sigma$ uncertainties are $1.05^{+0.04}_{-0.04}$, $1.10^{+0.04}_{-0.04}$, $1.21^{+0.05}_{-0.05}$, and $1.10^{+0.04}_{-0.03} \times 10^{-11}$ erg s$^{-1}$ cm$^{-2}$ for the September 1998, July 2020, October 2021, and May 2023 intrinsic profiles, respectively. The \lya\ flux from the October 2021 visit, the visit with the detected AU Mic b \lya\ transit, differs most from the other three visits, by $2-3-\sigma$.

\section{AU Mic c's High-Energy Environment} \label{sec:highen}

We investigate the high-energy environment of the planet to evaluate the importance of atmospheric escape influence for this system. The mass loss rate and ionization state of AU Mic c's planetary outflow is strongly dependent on its high-energy irradiation -- AU Mic's X-ray and EUV output ($5 - 1170$ \AA), in particular.

AU Mic's quiescent X-ray luminosity is time variable, with measurements between $\sim3 - 5\times 10^{29}$ erg s$^{-1}$ \citep{1990A&A...228..403P,2000A&A...359.1035L,2023ApJ...951...33T}. The EUV is more uncertain because it is largely absorbed by the interstellar medium. The lack of observational constraints for the EUV --- from AU Mic and low mass stars, in general --- lead to large systematic uncertainties in the predicted high-energy irradiation of planets, spanning more than an order of magnitude between different inference methods \citep[e.g.,][]{2011A&A...532A...6S,2014SoPh..289..515F,2020ApJ...895....5P,2021ApJ...909...61T,2021ApJ...913...40D}.

For a quiet AU Mic, \cite{2015Icar..250..357C} report X-ray and EUV luminosities of $2.51 \times 10^{29}$ erg s$^{-1}$ and $5.75 \times 10^{28}$ erg s$^{-1}$ measured by {\it ROSAT} ($5.17 - 124$ \AA) and {\it EUVE} ($80 - 350$ \AA) observations, respectively. A lower limit on the quiet XUV irradiation (covering the larger range of $5 - 1170$ \AA) experienced by AU Mic c is therefore $8940$ erg s$^{-1}$ cm$^{-2}$.

\cite{2015Icar..250..357C} use the coronal model from \cite{2011A&A...532A...6S}, which is powered by coronal and transition region emission line measurements to determine the amount of emitting material at different temperatures (i.e., the differential emission measure method, DEM), to create synthetic X-ray and EUV spectra for a flaring AU Mic. They report a flaring X-ray luminosity of $1.11 \times 10^{30}$ erg s$^{-1}$ and a flaring EUV luminosity of $2.47 \times 10^{29}$ erg s$^{-1}$. AU Mic c could potentially receive $39300$ erg s$^{-1}$ cm$^{-2}$ in XUV irradiation during a flare. \cite{2020AJ....160..237F} found that, even for the mature M dwarf GJ 699 (Barnard's star), flares could increase the star's overall UV output by 4-5 times, which is agrees with the estimates from \cite{2015Icar..250..357C}.

\cite{2022AJ....164..110F} updated the AU Mic X-ray to EUV spectrum originally fit in \cite{2021ApJ...913...40D} by including additional H/He continuum contributions within their DEM method. Their quiet and flaring panchromatic spectra for AU Mic give XUV irradiation estimates of $6625$ and $10625$ erg s$^{-1}$ cm$^{-2}$, respectively, for AU Mic c. These values are about $2/3$ and $1/3$ the corresponding fluxes from \cite{2015Icar..250..357C}. \cite{2022AJ....164..110F} provide the spectral information necessary to integrate over wavelength to calculate the photoionization rate (only integrated fluxes are available from \citealt{2015Icar..250..357C}). We calculate a photoionization rate of $1.9\times 10^{-4}$ s$^{-1}$ with Equations 7 and 8 from \cite{2023AJ....166...77R}, which corresponds to a neutral hydrogen lifetime of $5300$s ($\sim 1.5$ hours). This shortens only slightly during a flare to $\sim 1.1$ hours.

\subsection{Planetary mass-loss rate} \label{sec:loss}

AU Mic c's mass loss may be within the energy-limited regime given its irradiation and potential bulk densities \citep{2016ApJ...816...34O}, meaning that the mass loss rate is proportional to the received XUV radiation which heats the atmosphere and drives the outflow. The energy-limited mass loss rate is shown by Equation~\ref{eq:loss}. It depends on the size ($R_{\text{p}}, K_{\text{eff}}$) and mass ($M_{\text{p}}$) of the planet and the heating efficiency of the planet's atmosphere ($\eta$).

\begin{equation} \label{eq:loss}
    \dot{M} = \eta \frac{\pi R^3_{\text{p}}F_{\text{XUV}}}{GM_{\text{p}}K_{\text{eff}}}
\end{equation}
where:
\begin{equation} \label{eq:Keff}
    K_{\text{eff}} = \frac{(a/R_{\text{p}} -1)^2 (2a/R_{\text{p}} +1)}{2(a/R_{\text{p}})^3}
\end{equation}

\cite{2023MNRAS.525..455D}, \cite{2022MNRAS.512.3060Z} and \cite{2021AJ....162..295C} report AU Mic c's mass as $14.2$, $22.2$ and $9.6$ \Mearth, respectively, measured using radial velocity observations of AU Mic. \cite{2021AA...649A.177M} estimate AU Mic c's mass from transit-timing variations as $13.6$ \Mearth. \cite{2024AA...689A.132M} jointly fit the transit and radial velocity data for AU Mic c (and b and d) to obtain a mass of $14.5$ \Mearth. These multiple mass measurements contribute to uncertainty in the mass loss rate. 

We estimated AU Mic c's mass loss rate as $(0.98 - 2.3)\eta \times 10^{10}$ g s$^{-1}$ given the planet's wide potential mass range ($22.2 - 9.6$ \Mearth), the quiescent XUV flux from \cite{2022AJ....164..110F}, and the properties listed in Table~\ref{tab:prop}. The instantaneous mass-loss rate increases proportionally to the XUV flux during a flare. Despite the large uncertainties in XUV irradiation, the mass loss rate for AU Mic c is incredibly large ($\gtrsim 10^{10}$ g s$^{-1}$).

\subsection{Ionization fraction}

While the hefty XUV irradiation experienced by AU Mic c likely drives significant mass loss as derived in Section~\ref{sec:loss}, it also photoionizes neutral hydrogen within the exosphere. The fraction of neutral to ionized hydrogen determines whether the planet's exosphere is observable in transit at \lya.

We simulated AU Mic c's escaping hydrogen with the 1D hydrodynamic model {\tt p-winds} \citep{2022A&A...659A..62D}. This model assumes spherically symmetric and isotropic mass loss. {\tt p-winds} can take either an integrated ionizing flux or spectrum as an input to calculate the neutral/ionized hydrogen fractions; if the integrated flux is provided it is assumed to be monochromatic. For the \cite{2015Icar..250..357C} XUV estimates, we are limited to providing the integrated flux, whereas we are able to provide the DEM spectra from \cite{2022AJ....164..110F}. This loss of wavelength information, combined with the difference in flux, cause a drastic difference between the two XUV estimation methods. For this reason, we use the spectra from \cite{2022AJ....164..110F} and only discuss the ionization structure broadly and caution interpreting any further detail. We used the quiet and flaring irradiation from \cite{2022AJ....164..110F} and the intermediate $14.2$ \Mearth\ mass measurement from \cite{2023MNRAS.525..455D}, which gave planetary mass loss rates of $(1.5$ and $2.5) \times 10^{10}$ g s$^{-1}$, respectively. We assumed a planetary outflow temperature of $9000$ K, which is consistent with some measurements \citep[e.g.,][]{2018ApJ...855L..11O}.

\begin{figure}
\epsscale{1.15}
\plotone{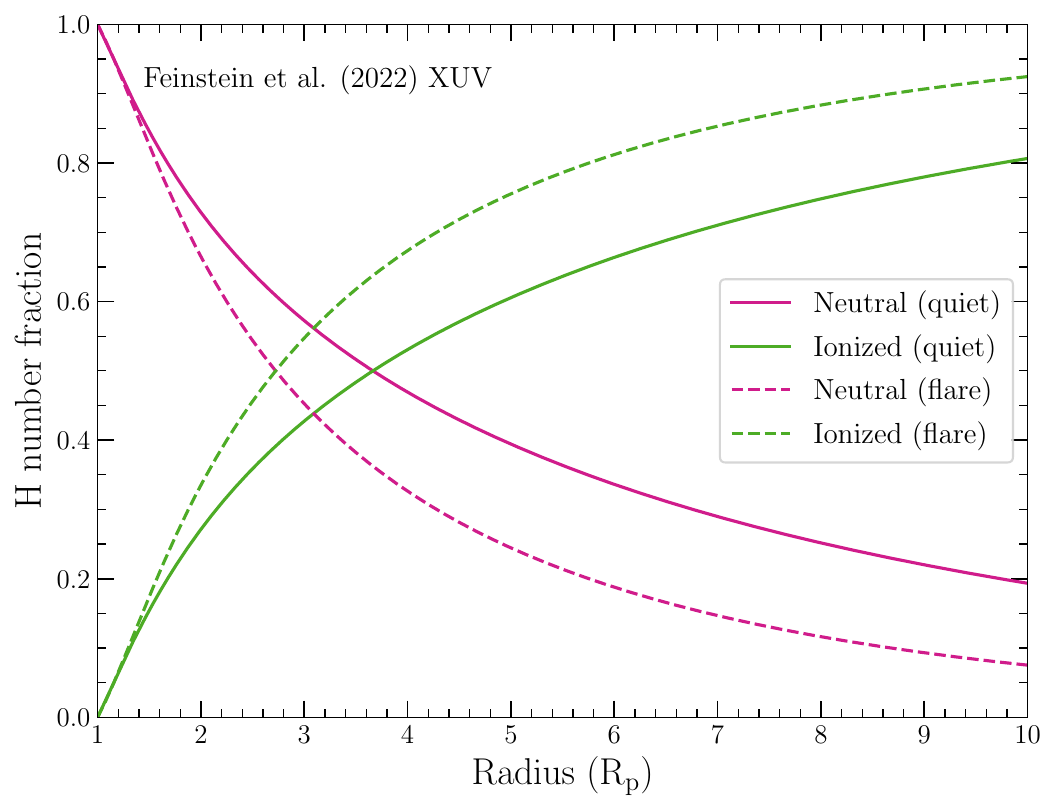}
\caption{The neutral and ionized hydrogen fractions of AU Mic c's exosphere as modeled by {\tt p-winds} using the XUV irradiation estimate from \cite{2022AJ....164..110F}. The solid lines represent the planet receiving quiescent irradiation, while the dashed lines represent flare irradiation. \label{fig:h_frac}}
\end{figure}

We compare the exosphere's neutral and ionized hydrogen fractions during quiescence and a flare in Figure~\ref{fig:h_frac}. The presence of neutral hydrogen does not change drastically given our flaring flux estimates, consistent with the small change in photoionization rate at AU Mic c. During a flare, the neutral-to-ionized reaches parity at about 3 $R_p$, compared to about 4 $R_p$ during quiescence.

The impact on the detectability of AU Mic c's exosphere is unclear. During a flare, the neutral hydrogen fraction at $8-10$ planetary radii is decreased by a factor of two, which will decrease the optical depth by the same factor \citep[e.g.][]{2023MNRAS.518.4357O}. This could have an impact on whether the material is observable at \lya\ since it is large planetary radii to which these observations are sensitive. Detailed modeling of stellar radiation and wind interactions --- which are important at large planetary radii --- is required. Finally, we caution that we are using XUV irradiation from a flaring template that cannot be straightforwardly scaled, not the XUV observed or modeled simultaneously with the flare in our May 2023 data.

\section{Discussion} \label{sec:discuss}

We do not detect drastic atmospheric escape of neutral hydrogen or other species within our one transit of AU Mic c with {\it HST}/STIS. This is in spite of the large amount of photoevaporation-inducing flux (F$_{\text{XUV}} = 8940$ erg s$^{-1}$ cm$^{-2}$) the planet is receiving. Our estimated mass loss rate of $(1.32 - 3.05)\eta \times 10^{10}$ g s$^{-1}$ is similar in magnitude to those for planets with detected atmospheric escape: $2.2\eta \times10^{10}$ g s$^{-1}$ for Gl 436 b \citep{2016A&A...591A.121B}, $1\eta \times10^{10}$ g s$^{-1}$ for GJ 3470 b \citep{2018A&A...620A.147B}, and $\sim30\eta \times 10^{10}$ g s$^{-1}$ for AU Mic b \citep{2023AJ....166...77R}. This leads us to the question: {\it why are we not detecting atmospheric escape off of AU Mic c?}

\subsection{Comparative exoplanetology between young atmospheric escape candidates}

\cite{2021AJ....162..116R}, within the context of K2-25 b, provide an extensive review of reasons why the atmospheric escape off of a young hot Neptune may not be detected. As K2-25 b also has a large predicted mass loss rate ($10.6\eta \times 10^{10}$ g s$^{-1}$), they posit that quick and sustained photoionization of the planet's escaping hydrogen prevents detection of atmospheric escape in \lya\ transmission \citep{2023MNRAS.518.4357O}. The same could be true for AU Mic c: the planet is losing neutral hydrogen, but the material is ionized before it is accelerated to observable speeds (i.e., the \lya\ wings). Our use of {\tt p-winds} to model AU Mic c's outflow indicates that this could be the case. If that is true, then one would expect AU Mic's closest planet, AU Mic b, to also experience too much photoionizing radiation to be detected. In fact, AU Mic b's quiescent XUV irradiation of $26000$ erg s$^{-1}$ cm$^{-2}$ (from the same luminosities used in Section~\ref{sec:highen}) is markedly higher than the $8763 \pm 1049$~erg~s$^{-1}$~cm$^{-2}$ estimated for K2-25 b and the $8940$ erg s$^{-1}$~cm$^{-2}$ experienced by AU Mic c. Despite this, neutral hydrogen has been detected escaping AU Mic b, albeit temporarily \citep{2023AJ....166...77R}. This work, when put into the context of other detections and non-detections, shows that photoionization's impact on detectability of atmospheric escape is not straightforward (i.e., there may be other processes that overpower photoionization to dominate on transit timescales, like stellar wind interactions).

Stellar wind interactions with escaping planetary material could render that material undetectable in transmission \citep{2020MNRAS.498L..53C,2022ApJ...934..189C}. \cite{2023AJ....166...77R} discuss the power of stellar wind ram pressure, even with intermediate strength, at shaping planetary exospheres showed by the 3D hydrodynamic simulations performed by \cite{2019ApJ...873...89M} and suggest this as the cause for the variable \lya\ transit of AU Mic b. AU Mic c is farther away from its host than AU Mic b and is experiencing a weaker stellar wind environment. AU Mic c's escaping planetary material, if there is any, would more easily spread out along the planet's orbit and create an observable transiting tail.

Then, does AU Mic c even have neutral hydrogen to lose? It is not likely that AU Mic c lost all of its neutral hydrogen already while the more highly irradiated AU Mic b has retained its neutral hydrogen. However, it is not clear if AU Mic b \& c were born with similar compositions. AU Mic b has a bulk density around $1 - 2$ g cm$^{-3}$, consistent with a small core surrounded by a large H/He envelope. AU Mic c, on the other hand, has a larger potential range in density, and therefore composition, because of the uncertainty in its mass. The highest mass measurement of $22.2$ \Mearth\ gives a bulk density of $3.66$ g cm $^{-3}$ which, if true, would mean that AU Mic c has a considerably smaller atmosphere than previously thought \citep{2022MNRAS.512.3060Z}. The bulk density is even larger ($7.62$ g cm $^{-3}$) if the \cite{2022MNRAS.512.3060Z} mass measurement is combined with the smaller radius measurement from \cite{2023AJ....166..232W}. This would explain our non-detection of hydrogen escape. The other three mass measurements characterize AU Mic c as having a similar bulk density and composition to AU Mic b. We need a conclusive measurement of AU Mic c's mass to distinguish between these scenarios.

Planets like K2-25 b, AU Mic b \& AU Mic c are likely the precursors to smaller sub-Neptunes and super-Earths, which are the most common exoplanets in the Milky Way. The comparison between these young planets shows the importance of investigating the presence and behavior of atmospheric escape in young hot Neptune systems. Our work on AU Mic c advises researchers in exoplanet atmosphere simulations to take into account a planet's environmental influences. Explicitly, photoionization {\it and} stellar wind need to be modeled together. This work also shows the importance of constraining exoplanet composition.

\begin{figure*}[ht!]
\epsscale{1.15}
\plotone{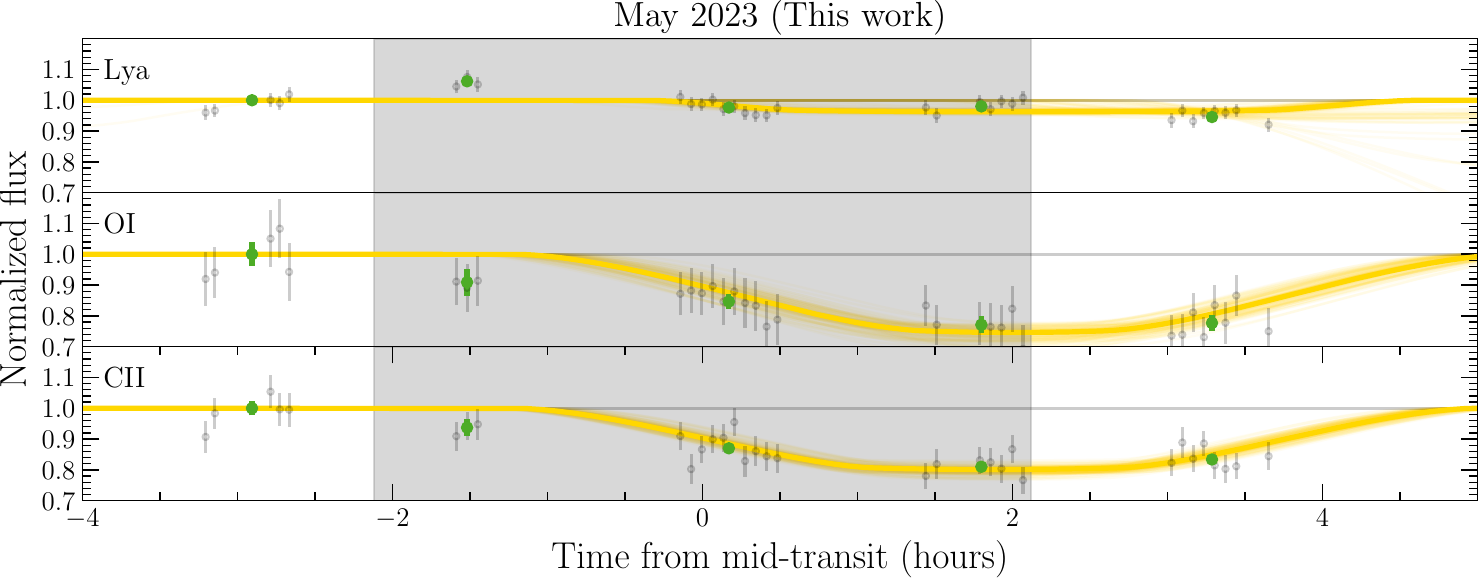}
\caption{AU Mic \lya, \ion{O}{1}, and \ion{C}{2} light curves observed by {\it HST}/STIS in May 2023 around the transit of AU Mic c. The gray region indicates the white-light transit duration of AU Mic c. Sample fits to the AU Mic c transit are shown in faded yellow with bright gold representing the best-fit light curves. \label{fig:lyac2o1_lcurve}}
\end{figure*}

\subsection{The challenge of interpreting FUV light curves}

The ultimate conclusion of this work is that the {\it HST}/STIS transmission spectra of AU Mic c's transit show {\it no} absorption due to escaping planetary material. We come to this conclusion based on our comparison with previous AU Mic data. However, it is important to note that if we only had the one transit of AU Mic c and no archival data, our results would change: the observations of AU Mic c mimic a convincing metal line transit.

We fit all of the AU Mic c emission line light curves (the right columns of Figures~\ref{fig:lcurve} and~\ref{fig:other_lcurve}) assuming the presence of a transiting exosphere. We used the {\tt batman} software to model the light curves using the transit properties of a circular opaque disk: mid-transit time ($t_0$), planetary exosphere size ($R_{\text{p}}/R_{\star}$), orbital period ($P_{\text{p}}$), semi-major axis ($a$), eccentricity ($e$), longitude of periastron ($\omega$), inclination ($i$), and limb-darkening coefficients ($u$). This model is limited by the assumption of a symmetrical circular transiting object, which doesn't allow us to fit asymmetric transit shapes -- as would be expected from a transiting tail of escaping material. We fit this model to our AU Mic c sub-exposure light curves using the MCMC algorithm {\tt emcee} \citep{2013PASP..125..306F}. We fit for the mid-transit time ($t_0=[-10,10]$ hours), exosphere size ($R_{\text{p}}/R_{\star} > 0$) and quadratic limb darkening coefficients ($u > 0$) using uniform priors. The remaining transit parameters are fixed to those determined in \cite{2023AJ....166..232W}.

We ran the algorithm with $9,000$ steps, $90$ walkers, and a $900$ burn-in. The significance of every emission line light curve's best-fit transit depth (Equation~\ref{eq:depth_sig}) was then calculated and compared. The transit depth significance for the \lya, \ion{O}{1}, and \ion{C}{2} light curves were all above $3$, with \ion{C}{2} being the most significant at $\frac{\delta}{\sigma_{\delta}} = 15.0$. The remaining emission line light curves had transit depth significance below $1$.

\begin{equation} \label{eq:depth_sig}
    \frac{\delta}{\sigma_{\delta}} = \frac{R_{\text{p}}/R_{\star}}{2\sigma_{R_{\text{p}}/R_{\star}}}
\end{equation}

Essentially, we can fit significant planetary transits to the \lya\, \ion{O}{1}, and \ion{C}{2} light curves shown in Figure~\ref{fig:lyac2o1_lcurve}. As a reminder, {\it we ruled out any actual transit of AU Mic c} in Section~\ref{sec:pl_search} where we determined that the flux data during our AU Mic c observation did not significantly differ from the archival out-of-transit AU Mic flux behavior (for all emission lines considered in this work). If we did not have the wealth of FUV data that we do for AU Mic to compare to, the single visit to AU Mic c would present a false detection of escaping metal material. This motivates current and future programs investigating the ultraviolet variability of young cool stars.

\section{Summary and conclusions} \label{sec:summary}

We embarked on a journey to witness AU Mic c's atmosphere actively escaping. The pursuit of observing and characterizing atmospheric escape on young planets is motivated by the strong theoretical link between atmospheric escape and the overall evolution of the exoplanet population, ultimately producing the current picture of exoplanet demographics. Here are our main conclusions:
\begin{itemize}
    \item[1.] We presented new {\it HST}/STIS far-ultraviolet data taken around the transit of AU Mic c (May 2023) and compared it to three previous visits of AU Mic --- two visits corresponding to the white-light transit of AU Mic b (July 2020 and October 2021), and one out-of-transit visit (September 1998).
    \item[2.] We detected 6 flares in the \ion{Si}{4} and \ion{C}{4} emission lines around the transit of AU Mic c (May 2023). In total, 18 flares were detected across all four AU Mic visits. We modeled each of the May 2023 flares using a modified version of {\tt cos\_flares}. The longest flare had an equivalent duration of $15$ minutes. Only one of the May 2023 flares corresponded to a \lya\ flare, which masked a $10$\% transit for $8.5$ minutes. If flares were the sole source of variability, a large neutral hydrogen exosphere should still have been detectable within the data.
    \item[3.] There are $10$\% variations in \lya\, \ion{O}{1}, and \ion{C}{2} flux over the course of the AU Mic c transit that we reported as stellar in nature due to the similar behavior observed in the archival AU Mic data, the clear influence of flares, and the lack of velocity-dependent flux decrements.
    \item[4.] We estimated AU Mic c's quiescent XUV irradiation to be about $6625$ erg s$^{-1}$ cm$^{-2}$. Its energy-limited mass loss rate, given its potential mass range of $22.2 - 9.6$ \Mearth, is then $(0.98-2.3)\eta \times 10^{10}$ g s$^{-1}$. The quiescent photoionization rate at AU Mic c, derived from the EUV DEM from \cite{2022AJ....164..110F}, is about $1.5$ hours.
\end{itemize}

\section*{Acknowledgements}
\vspace{2mm}
\begin{center}
    In loving memory of Judy Rockcliffe and Eileen Tavares. Two caring and wonderful grandmothers. Judy, a huge beach bum, and Eileen, who always made everyone smile.
\end{center}

This research is based on observations made with the NASA/ESA {\it Hubble Space Telescope} obtained from the Space Telescope Science Institute, which is operated by the Association of Universities for Research in Astronomy, Inc., under NASA contract NAS 5–26555. These observations are associated with HST-GO-17219. Support for program HST-GO-17219 was provided by NASA through a grant from the STScI.

This research was carried out in part at the Jet Propulsion Laboratory, California Institute of Technology, under a contract with the National Aeronautics and Space Administration (80NM0018D0004). ADF acknowledges funding from NASA through the NASA Hubble Fellowship grant HST-HF2-51530.001-A awarded by STScI.

Some/all of the data presented in this article were obtained from the Mikulski Archive for Space Telescopes (MAST) at the Space Telescope Science Institute. The five {\it HST}/STIS exposures completed in May 2023 can be accessed via \dataset[DOI: 10.17909/qz7b-bf83]{https://doi.org/10.17909/qz7b-bf83}. The twelve {\it HST}/STIS exposures completed in July 2020 and October 2021 surrounding the transit of AU Mic b can be accessed via \dataset[DOI: 10.17909/5hz4-vx12]{https://doi.org/10.17909/5hz4-vx12}. Finally, the four {\it HST}/STIS exposures completed in September 1998 of AU Mic are available at \dataset[DOI: 10.17909/ekgk-8d20]{https://doi.org/10.17909/ekgk-8d20}.

%

\facilities{HST/STIS \citep{1998PASP..110.1183W}}


\software{{\tt astropy} \citep{astropy:2013, astropy:2018, astropy:2022}, {\tt batman} \citep{2015PASP..127.1161K}, {\tt cos\_flares} \citep{2022AJ....164..110F}, {\tt emcee} \citep{2013PASP..125..306F}, {\tt lightkurve} \citep{2018ascl.soft12013L}, {\tt lyapy} \citep{2022zndo...6949067Y}, {\tt matplotlib} \citep{hunter2007matplotlib}, {\tt numpy} \citep{harris2020array}, {\tt p-winds} \citep{2022A&A...659A..62D}, {\tt scipy} \citep{2020SciPy}, {\tt stistools} (\url{https://github.com/spacetelescope/stistools})}


\appendix
\restartappendixnumbering
\section{Additional light curve information}
\subsection{Supplemental Far-Ultraviolet emission line behavior} \label{ap:lines}
\begin{figure*}
\epsscale{1.15}
\plotone{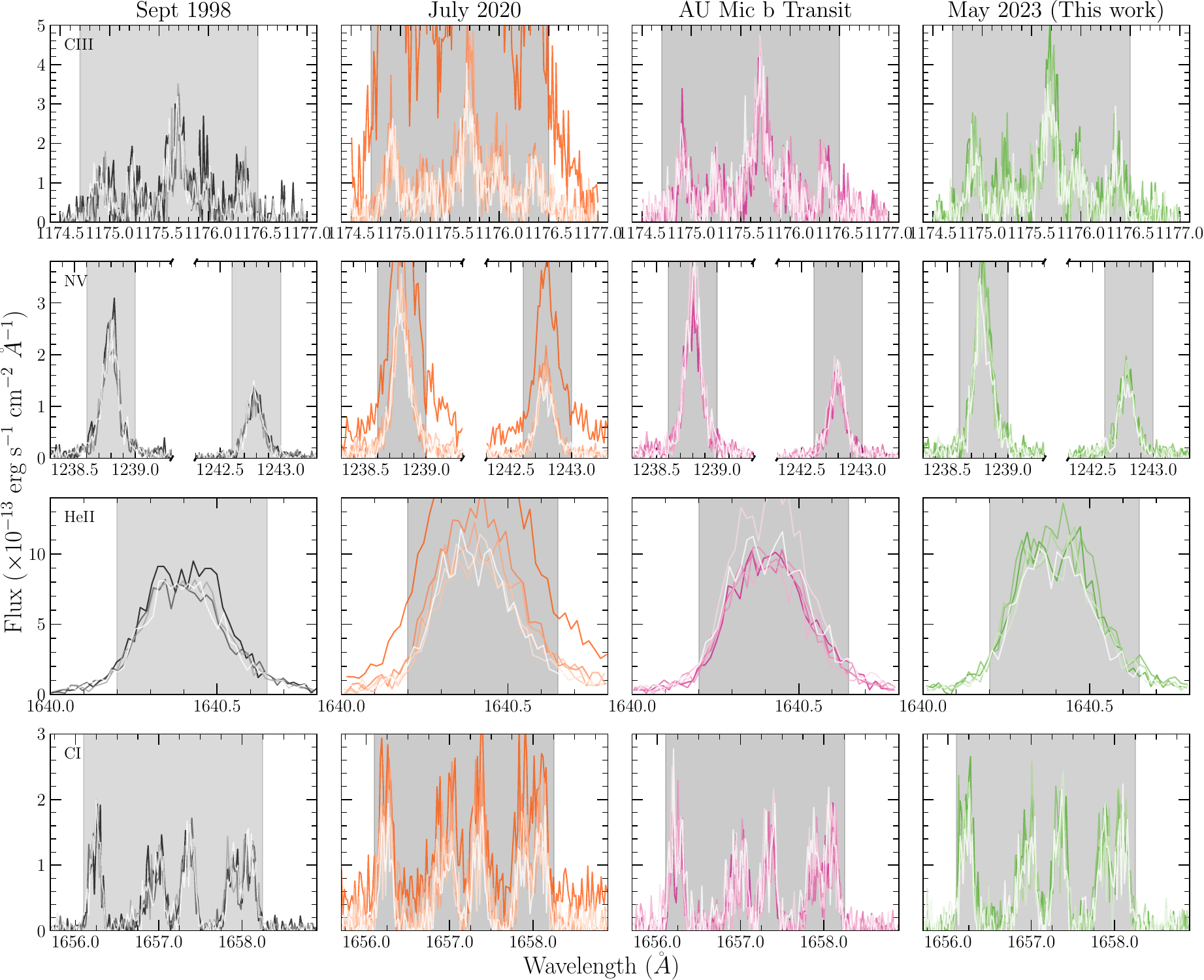}
\caption{AU Mic spectra observed by {\it HST}/STIS on September 1998 (left; no transit), July 2020 (second-from-left; undetected AU Mic b transit), and October 2021 (middle; detected AU Mic b transit) and May 2023 (right; during AU Mic c transit). \ion{C}{3}, \ion{N}{5}, \ion{He}{2}, and \ion{C}{1} are shown. The gray regions show the wavelength ranges integrated over to obtain the light curves in Figure~\ref{fig:other_lcurve}. Each line color corresponds to same orbit from Figure~\ref{fig:lyaspec}. \label{fig:other_spec}}
\end{figure*}

\begin{figure*}[ht!]
\epsscale{1.15}
\plotone{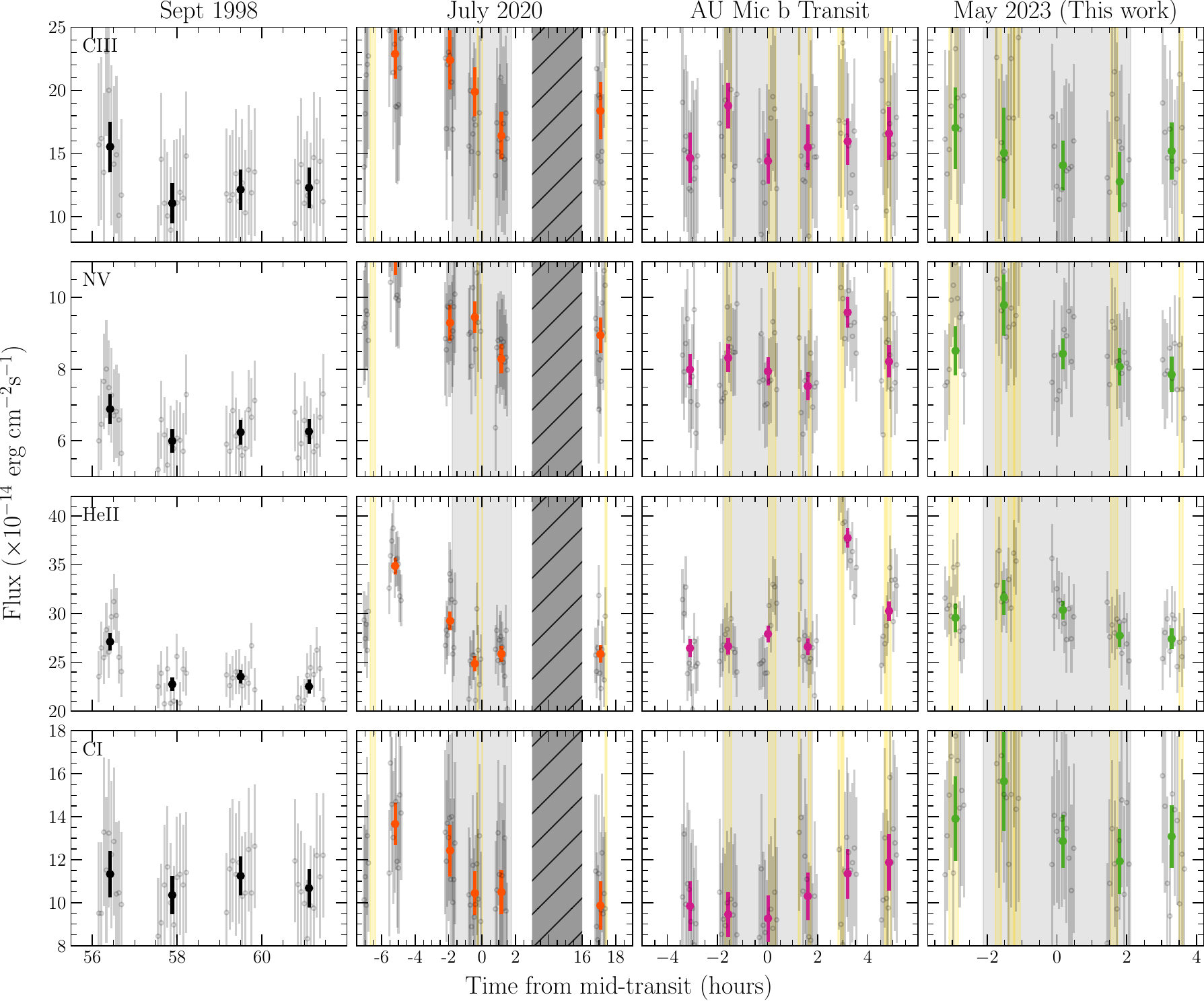}
\caption{AU Mic spectra observed by {\it HST}/STIS on September 1998 (left; no transit), July 2020 (second-from-left; undetected AU Mic b transit), and October 2021 (middle; detected AU Mic b transit) and May 2023 (right; during AU Mic c transit). \ion{C}{3}, \ion{N}{5}, \ion{He}{2}, and \ion{C}{1} are shown. The gray regions indicate the white-light transit durations of AU Mic b and c. Flares are highlighted in gold. \label{fig:other_lcurve}}
\end{figure*}

All FUV emission lines analyzed but not referenced in this work (\ion{C}{3}, \ion{N}{5}, \ion{He}{2}, and \ion{C}{1}) are included here as supplemental information (Figures~\ref{fig:other_spec} and~\ref{fig:other_lcurve}).

\subsection{Far-Ultraviolet continuum} \label{ap:cont}
\begin{figure}
\epsscale{1.15}
\plotone{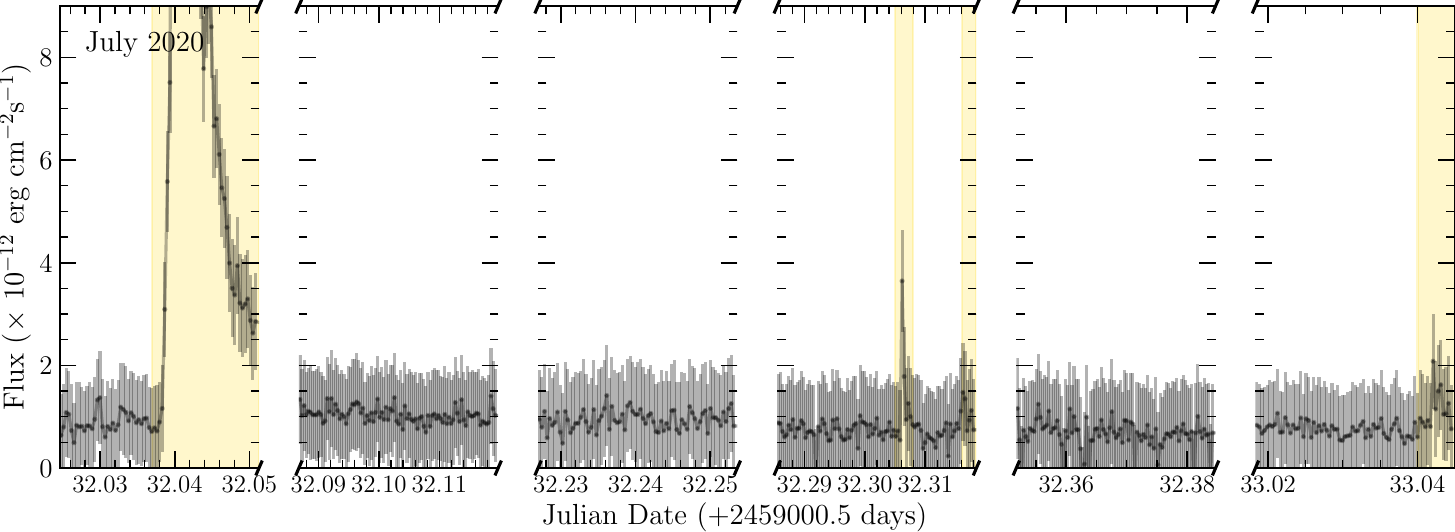}
\plotone{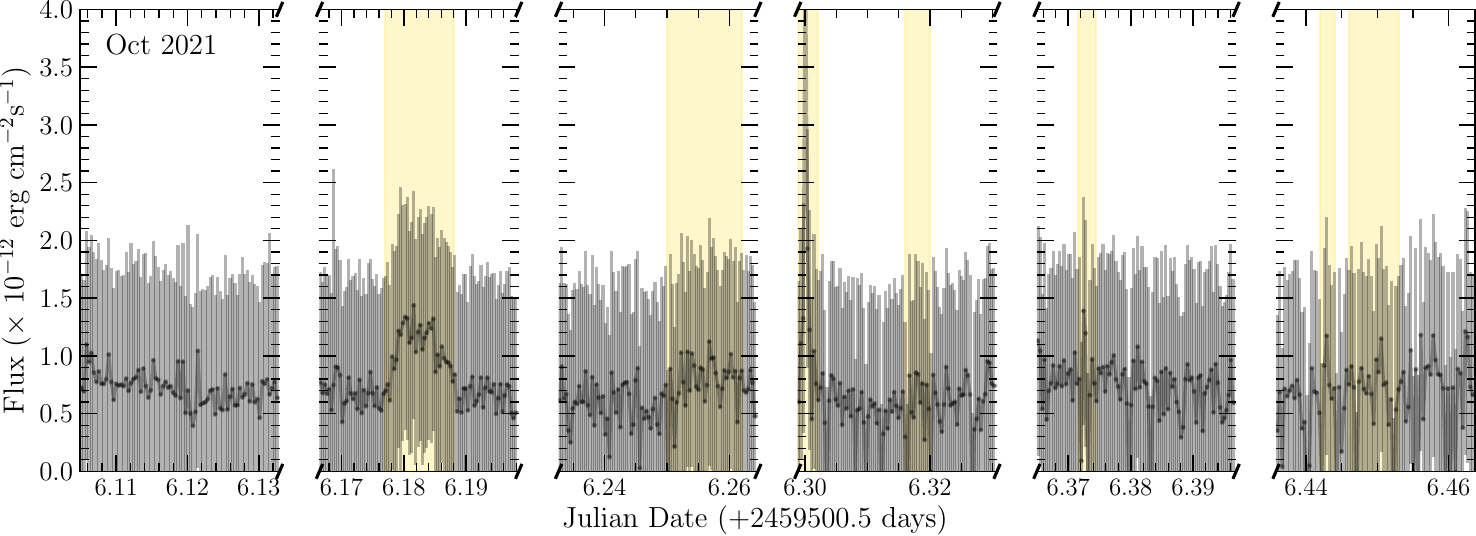}
\plotone{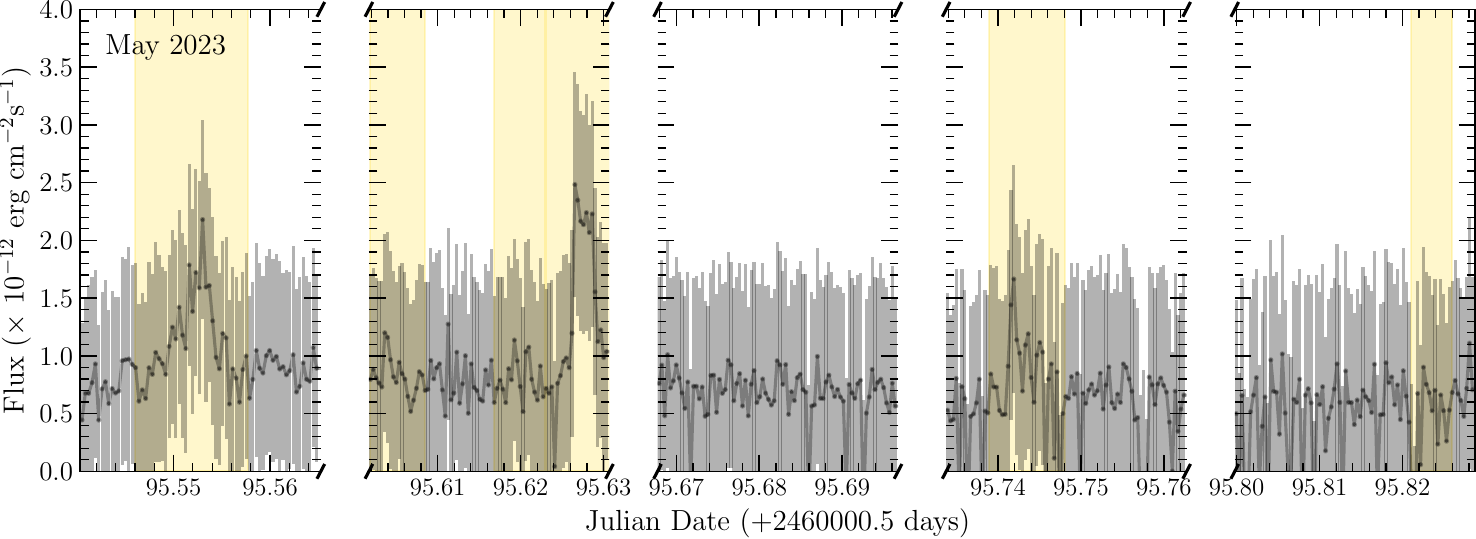}
\caption{AU Mic's FUV emission over time during the July 2020 (top; AU Mic b transit), October 2021 (middle; AU Mic b transit), and May 2023 (bottom; AU Mic c transit) visits. Flares identified in Section~\ref{sec:flares} are highlighted in gold.}
\label{fig:cont_flares}
\end{figure}

We defined the FUV continuum using the same wavelength regions as \cite{2018ApJ...867...71L}, although we removed their bluest wavelength bin due to the large noise in that region of the STIS/E140M spectra: $1201.71 - 1212.16$, $1219.18 - 1274.04$, $1329.25 - 1354.49$, $1356.71 - 1357.59$, $1359.51 - 1428.90$ \AA. These regions were summed to create the light curves in Figure~\ref{fig:cont_flares}.

\begin{thebibliography}{}
\expandafter\ifx\csname natexlab\endcsname\relax\def\natexlab#1{#1}\fi
\providecommand{\url}[1]{\href{#1}{#1}}
\providecommand{\dodoi}[1]{doi:~\href{http://doi.org/#1}{\nolinkurl{#1}}}
\providecommand{\doeprint}[1]{\href{http://ascl.net/#1}{\nolinkurl{http://ascl.net/#1}}}
\providecommand{\doarXiv}[1]{\href{https://arxiv.org/abs/#1}{\nolinkurl{https://arxiv.org/abs/#1}}}

\bibitem[{{Astropy Collaboration} {et~al.}(2013){Astropy Collaboration},
  {Robitaille}, {Tollerud}, {Greenfield}, {Droettboom}, {Bray}, {Aldcroft},
  {Davis}, {Ginsburg}, {Price-Whelan}, {Kerzendorf}, {Conley}, {Crighton},
  {Barbary}, {Muna}, {Ferguson}, {Grollier}, {Parikh}, {Nair}, {Unther},
  {Deil}, {Woillez}, {Conseil}, {Kramer}, {Turner}, {Singer}, {Fox}, {Weaver},
  {Zabalza}, {Edwards}, {Azalee Bostroem}, {Burke}, {Casey}, {Crawford},
  {Dencheva}, {Ely}, {Jenness}, {Labrie}, {Lim}, {Pierfederici}, {Pontzen},
  {Ptak}, {Refsdal}, {Servillat}, \& {Streicher}}]{astropy:2013}
{Astropy Collaboration}, {Robitaille}, T.~P., {Tollerud}, E.~J., {et~al.} 2013,
  \aap, 558, A33, \dodoi{10.1051/0004-6361/201322068}

\bibitem[{{Astropy Collaboration} {et~al.}(2018){Astropy Collaboration},
  {Price-Whelan}, {Sip{\H{o}}cz}, {G{\"u}nther}, {Lim}, {Crawford}, {Conseil},
  {Shupe}, {Craig}, {Dencheva}, {Ginsburg}, {VanderPlas}, {Bradley},
  {P{\'e}rez-Su{\'a}rez}, {de Val-Borro}, {Aldcroft}, {Cruz}, {Robitaille},
  {Tollerud}, {Ardelean}, {Babej}, {Bach}, {Bachetti}, {Bakanov}, {Bamford},
  {Barentsen}, {Barmby}, {Baumbach}, {Berry}, {Biscani}, {Boquien}, {Bostroem},
  {Bouma}, {Brammer}, {Bray}, {Breytenbach}, {Buddelmeijer}, {Burke},
  {Calderone}, {Cano Rodr{\'\i}guez}, {Cara}, {Cardoso}, {Cheedella}, {Copin},
  {Corrales}, {Crichton}, {D'Avella}, {Deil}, {Depagne}, {Dietrich}, {Donath},
  {Droettboom}, {Earl}, {Erben}, {Fabbro}, {Ferreira}, {Finethy}, {Fox},
  {Garrison}, {Gibbons}, {Goldstein}, {Gommers}, {Greco}, {Greenfield},
  {Groener}, {Grollier}, {Hagen}, {Hirst}, {Homeier}, {Horton}, {Hosseinzadeh},
  {Hu}, {Hunkeler}, {Ivezi{\'c}}, {Jain}, {Jenness}, {Kanarek}, {Kendrew},
  {Kern}, {Kerzendorf}, {Khvalko}, {King}, {Kirkby}, {Kulkarni}, {Kumar},
  {Lee}, {Lenz}, {Littlefair}, {Ma}, {Macleod}, {Mastropietro}, {McCully},
  {Montagnac}, {Morris}, {Mueller}, {Mumford}, {Muna}, {Murphy}, {Nelson},
  {Nguyen}, {Ninan}, {N{\"o}the}, {Ogaz}, {Oh}, {Parejko}, {Parley}, {Pascual},
  {Patil}, {Patil}, {Plunkett}, {Prochaska}, {Rastogi}, {Reddy Janga},
  {Sabater}, {Sakurikar}, {Seifert}, {Sherbert}, {Sherwood-Taylor}, {Shih},
  {Sick}, {Silbiger}, {Singanamalla}, {Singer}, {Sladen}, {Sooley},
  {Sornarajah}, {Streicher}, {Teuben}, {Thomas}, {Tremblay}, {Turner},
  {Terr{\'o}n}, {van Kerkwijk}, {de la Vega}, {Watkins}, {Weaver}, {Whitmore},
  {Woillez}, {Zabalza}, \& {Astropy Contributors}}]{astropy:2018}
{Astropy Collaboration}, {Price-Whelan}, A.~M., {Sip{\H{o}}cz}, B.~M., {et~al.}
  2018, \aj, 156, 123, \dodoi{10.3847/1538-3881/aabc4f}

\bibitem[{{Astropy Collaboration} {et~al.}(2022){Astropy Collaboration},
  {Price-Whelan}, {Lim}, {Earl}, {Starkman}, {Bradley}, {Shupe}, {Patil},
  {Corrales}, {Brasseur}, {N{\"o}the}, {Donath}, {Tollerud}, {Morris},
  {Ginsburg}, {Vaher}, {Weaver}, {Tocknell}, {Jamieson}, {van Kerkwijk},
  {Robitaille}, {Merry}, {Bachetti}, {G{\"u}nther}, {Aldcroft},
  {Alvarado-Montes}, {Archibald}, {B{\'o}di}, {Bapat}, {Barentsen},
  {Baz{\'a}n}, {Biswas}, {Boquien}, {Burke}, {Cara}, {Cara}, {Conroy},
  {Conseil}, {Craig}, {Cross}, {Cruz}, {D'Eugenio}, {Dencheva}, {Devillepoix},
  {Dietrich}, {Eigenbrot}, {Erben}, {Ferreira}, {Foreman-Mackey}, {Fox},
  {Freij}, {Garg}, {Geda}, {Glattly}, {Gondhalekar}, {Gordon}, {Grant},
  {Greenfield}, {Groener}, {Guest}, {Gurovich}, {Handberg}, {Hart},
  {Hatfield-Dodds}, {Homeier}, {Hosseinzadeh}, {Jenness}, {Jones}, {Joseph},
  {Kalmbach}, {Karamehmetoglu}, {Ka{\l}uszy{\'n}ski}, {Kelley}, {Kern},
  {Kerzendorf}, {Koch}, {Kulumani}, {Lee}, {Ly}, {Ma}, {MacBride}, {Maljaars},
  {Muna}, {Murphy}, {Norman}, {O'Steen}, {Oman}, {Pacifici}, {Pascual},
  {Pascual-Granado}, {Patil}, {Perren}, {Pickering}, {Rastogi}, {Roulston},
  {Ryan}, {Rykoff}, {Sabater}, {Sakurikar}, {Salgado}, {Sanghi}, {Saunders},
  {Savchenko}, {Schwardt}, {Seifert-Eckert}, {Shih}, {Jain}, {Shukla}, {Sick},
  {Simpson}, {Singanamalla}, {Singer}, {Singhal}, {Sinha}, {Sip{\H{o}}cz},
  {Spitler}, {Stansby}, {Streicher}, {{\v{S}}umak}, {Swinbank}, {Taranu},
  {Tewary}, {Tremblay}, {de Val-Borro}, {Van Kooten}, {Vasovi{\'c}}, {Verma},
  {de Miranda Cardoso}, {Williams}, {Wilson}, {Winkel}, {Wood-Vasey}, {Xue},
  {Yoachim}, {Zhang}, {Zonca}, \& {Astropy Project
  Contributors}}]{astropy:2022}
{Astropy Collaboration}, {Price-Whelan}, A.~M., {Lim}, P.~L., {et~al.} 2022,
  \apj, 935, 167, \dodoi{10.3847/1538-4357/ac7c74}

\bibitem[{{Ben-Jaffel} {et~al.}(2022){Ben-Jaffel}, {Ballester}, {Mu{\~n}oz},
  {Lavvas}, {Sing}, {Sanz-Forcada}, {Cohen}, {Kataria}, {Henry}, {Buchhave},
  {Mikal-Evans}, {Wakeford}, \& {L{\'o}pez-Morales}}]{2022NatAs...6..141B}
{Ben-Jaffel}, L., {Ballester}, G.~E., {Mu{\~n}oz}, A.~G., {et~al.} 2022, Nature
  Astronomy, 6, 141, \dodoi{10.1038/s41550-021-01505-x}

\bibitem[{{Boldog} {et~al.}(2025){Boldog}, {Szab{\'o}}, {Kriskovics},
  {Borsato}, {Gandolfi}, {Lendl}, {G{\"u}nther}, {Heitzmann}, {Wilson},
  {Brandeker}, {Garai}, {Alibert}, {Alonso}, {B{\'a}rczy}, {Barrado Navascues},
  {Barros}, {Baumjohann}, {Benz}, {Billot}, {Broeg}, {Collier Cameron},
  {Correia}, {Csizmadia}, {Cubillos}, {Davies}, {Deleuil}, {Deline},
  {Demangeon}, {Demory}, {Derekas}, {Edwards}, {Egger}, {Ehrenreich},
  {Erikson}, {Fortier}, {Fossati}, {Fridlund}, {Gazeas}, {Gillon}, {G{\"u}del},
  {Guterman}, {Helling}, {Isaak}, {Kiss}, {Kopp}, {Korth}, {Lam}, {Laskar},
  {Lecavelier des Etangs}, {Luntzer}, {Magrin}, {Mantovan}, {Marafatto},
  {Maxted}, {Mer{\'\i}n}, {Mordasini}, {Munari}, {Nascimbeni}, {Olofsson},
  {Ottensamer}, {Pagano}, {Pall{\'e}}, {Peter}, {Piazza}, {Piotto}, {Pollacco},
  {Poppenhaeger}, {Queloz}, {Ragazzoni}, {Rando}, {Rauer}, {Ribas}, {Rieder},
  {Santos}, {Scandariato}, {S{\'e}gransan}, {Simon}, {Smith}, {Sousa},
  {Southworth}, {Stalport}, {Sulis}, {Udry}, {Ulmer-Moll}, {Van Grootel},
  {Venturini}, {Villaver}, {Walton}, \& {Zingales}}]{2025A&A...694A.137B}
{Boldog}, {\'A}., {Szab{\'o}}, G.~M., {Kriskovics}, L., {et~al.} 2025, \aap,
  694, A137, \dodoi{10.1051/0004-6361/202452699}

\bibitem[{{Bourrier} \& {Lecavelier des Etangs}(2013)}]{2013A&A...557A.124B}
{Bourrier}, V., \& {Lecavelier des Etangs}, A. 2013, \aap, 557, A124,
  \dodoi{10.1051/0004-6361/201321551}

\bibitem[{{Bourrier} {et~al.}(2016){Bourrier}, {Lecavelier des Etangs},
  {Ehrenreich}, {Tanaka}, \& {Vidotto}}]{2016A&A...591A.121B}
{Bourrier}, V., {Lecavelier des Etangs}, A., {Ehrenreich}, D., {Tanaka}, Y.~A.,
  \& {Vidotto}, A.~A. 2016, \aap, 591, A121,
  \dodoi{10.1051/0004-6361/201628362}

\bibitem[{{Bourrier} {et~al.}(2018){Bourrier}, {Lecavelier des Etangs},
  {Ehrenreich}, {Sanz-Forcada}, {Allart}, {Ballester}, {Buchhave}, {Cohen},
  {Deming}, {Evans}, {Garc{\'\i}a Mu{\~n}oz}, {Henry}, {Kataria}, {Lavvas},
  {Lewis}, {L{\'o}pez-Morales}, {Marley}, {Sing}, \&
  {Wakeford}}]{2018A&A...620A.147B}
{Bourrier}, V., {Lecavelier des Etangs}, A., {Ehrenreich}, D., {et~al.} 2018,
  \aap, 620, A147, \dodoi{10.1051/0004-6361/201833675}

\bibitem[{{Bourrier} {et~al.}(2022){Bourrier}, {Zapatero Osorio}, {Allart},
  {Attia}, {Cretignier}, {Dumusque}, {Lovis}, {Adibekyan}, {Borsa}, {Figueira},
  {Gonz{\'a}lez Hern{\'a}ndez}, {Mehner}, {Santos}, {Schmidt}, {Seidel},
  {Sozzetti}, {Alibert}, {Casasayas-Barris}, {Ehrenreich}, {Lo Curto},
  {Martins}, {Di Marcantonio}, {M{\'e}gevand}, {Nunes}, {Palle}, {Poretti}, \&
  {Sousa}}]{2022A&A...663A.160B}
{Bourrier}, V., {Zapatero Osorio}, M.~R., {Allart}, R., {et~al.} 2022, \aap,
  663, A160, \dodoi{10.1051/0004-6361/202142559}

\bibitem[{{Bourrier} {et~al.}(2023){Bourrier}, {Attia}, {Mallonn}, {Marret},
  {Lendl}, {Konig}, {Krenn}, {Cretignier}, {Allart}, {Henry}, {Bryant},
  {Leleu}, {Nielsen}, {Hebrard}, {Hara}, {Ehrenreich}, {Seidel}, {dos Santos},
  {Lovis}, {Bayliss}, {Cegla}, {Dumusque}, {Boisse}, {Boucher}, {Bouchy},
  {Pepe}, {Lavie}, {Rey Cerda}, {S{\'e}gransan}, {Udry}, \&
  {Vrignaud}}]{2023A&A...669A..63B}
{Bourrier}, V., {Attia}, O., {Mallonn}, M., {et~al.} 2023, \aap, 669, A63,
  \dodoi{10.1051/0004-6361/202245004}

\bibitem[{{Cale} {et~al.}(2021){Cale}, {Reefe}, {Plavchan}, {Tanner}, {Gaidos},
  {Gagn{\'e}}, {Gao}, {Kane}, {B{\'e}jar}, {Lodieu}, {Anglada-Escud{\'e}},
  {Ribas}, {Pall{\'e}}, {Quirrenbach}, {Amado}, {Reiners}, {Caballero}, {Rosa
  Zapatero Osorio}, {Dreizler}, {Howard}, {Fulton}, {Xuesong Wang}, {Collins},
  {El Mufti}, {Wittrock}, {Gilbert}, {Barclay}, {Klein}, {Martioli},
  {Wittenmyer}, {Wright}, {Addison}, {Hirano}, {Tamura}, {Kotani}, {Narita},
  {Vermilion}, {Lee}, {Geneser}, {Teske}, {Quinn}, {Latham}, {Esquerdo},
  {Calkins}, {Berlind}, {Zohrabi}, {Stibbards}, {Kotnana}, {Jenkins},
  {Twicken}, {Henze}, {Kidwell}, {Burke}, {Villase{\~n}or}, \&
  {Boyd}}]{2021AJ....162..295C}
{Cale}, B.~L., {Reefe}, M., {Plavchan}, P., {et~al.} 2021, \aj, 162, 295,
  \dodoi{10.3847/1538-3881/ac2c80}

\bibitem[{{Carolan} {et~al.}(2020){Carolan}, {Vidotto}, {Plavchan}, {Villarreal
  D'Angelo}, \& {Hazra}}]{2020MNRAS.498L..53C}
{Carolan}, S., {Vidotto}, A.~A., {Plavchan}, P., {Villarreal D'Angelo}, C., \&
  {Hazra}, G. 2020, \mnras, 498, L53, \dodoi{10.1093/mnrasl/slaa127}

\bibitem[{{Chadney} {et~al.}(2015){Chadney}, {Galand}, {Unruh}, {Koskinen}, \&
  {Sanz-Forcada}}]{2015Icar..250..357C}
{Chadney}, J.~M., {Galand}, M., {Unruh}, Y.~C., {Koskinen}, T.~T., \&
  {Sanz-Forcada}, J. 2015, \icarus, 250, 357,
  \dodoi{10.1016/j.icarus.2014.12.012}

\bibitem[{{Chen} \& {Rogers}(2016)}]{2016ApJ...831..180C}
{Chen}, H., \& {Rogers}, L.~A. 2016, \apj, 831, 180,
  \dodoi{10.3847/0004-637X/831/2/180}

\bibitem[{{Cohen} {et~al.}(2022){Cohen}, {Alvarado-G{\'o}mez}, {Drake},
  {Harbach}, {Garraffo}, \& {Fraschetti}}]{2022ApJ...934..189C}
{Cohen}, O., {Alvarado-G{\'o}mez}, J.~D., {Drake}, J.~J., {et~al.} 2022, \apj,
  934, 189, \dodoi{10.3847/1538-4357/ac78e4}

\bibitem[{{Del Zanna} {et~al.}(2021){Del Zanna}, {Dere}, {Young}, \&
  {Landi}}]{2021ApJ...909...38D}
{Del Zanna}, G., {Dere}, K.~P., {Young}, P.~R., \& {Landi}, E. 2021, \apj, 909,
  38, \dodoi{10.3847/1538-4357/abd8ce}

\bibitem[{{Dere} {et~al.}(1997){Dere}, {Landi}, {Mason}, {Monsignori Fossi}, \&
  {Young}}]{1997AAS..125..149D}
{Dere}, K.~P., {Landi}, E., {Mason}, H.~E., {Monsignori Fossi}, B.~C., \&
  {Young}, P.~R. 1997, \aaps, 125, 149, \dodoi{10.1051/aas:1997368}

\bibitem[{{Donati} {et~al.}(2023){Donati}, {Cristofari}, {Finociety}, {Klein},
  {Moutou}, {Gaidos}, {Cadieux}, {Artigau}, {Correia}, {Bou{\'e}}, {Cook},
  {Carmona}, {Lehmann}, {Bouvier}, {Martioli}, {Morin}, {Fouqu{\'e}},
  {Delfosse}, {Doyon}, {H{\'e}brard}, {Alencar}, {Laskar}, {Arnold}, {Petit},
  {K{\'o}sp{\'a}l}, {Vidotto}, {Folsom}, \&
  {collaboration}}]{2023MNRAS.525..455D}
{Donati}, J.~F., {Cristofari}, P.~I., {Finociety}, B., {et~al.} 2023, \mnras,
  525, 455, \dodoi{10.1093/mnras/stad1193}

\bibitem[{{Dos Santos} {et~al.}(2022){Dos Santos}, {Vidotto}, {Vissapragada},
  {Alam}, {Allart}, {Bourrier}, {Kirk}, {Seidel}, \&
  {Ehrenreich}}]{2022A&A...659A..62D}
{Dos Santos}, L.~A., {Vidotto}, A.~A., {Vissapragada}, S., {et~al.} 2022, \aap,
  659, A62, \dodoi{10.1051/0004-6361/202142038}

\bibitem[{{Dos Santos} {et~al.}(2023){Dos Santos}, {Garc{\'\i}a Mu{\~n}oz},
  {Sing}, {L{\'o}pez-Morales}, {Alam}, {Bourrier}, {Ehrenreich}, {Henry},
  {Lecavelier des Etangs}, {Mikal-Evans}, {Nikolov}, {Sanz-Forcada}, \&
  {Wakeford}}]{2023AJ....166...89D}
{Dos Santos}, L.~A., {Garc{\'\i}a Mu{\~n}oz}, A., {Sing}, D.~K., {et~al.} 2023,
  \aj, 166, 89, \dodoi{10.3847/1538-3881/ace445}

\bibitem[{{Duvvuri} {et~al.}(2021){Duvvuri}, {Sebastian Pineda},
  {Berta-Thompson}, {Brown}, {France}, {Kowalski}, {Redfield}, {Tilipman},
  {Vieytes}, {Wilson}, {Youngblood}, {Froning}, {Linsky}, {Parke Loyd},
  {Mauas}, {Miguel}, {Newton}, {Rugheimer}, \& {Christian
  Schneider}}]{2021ApJ...913...40D}
{Duvvuri}, G.~M., {Sebastian Pineda}, J., {Berta-Thompson}, Z.~K., {et~al.}
  2021, \apj, 913, 40, \dodoi{10.3847/1538-4357/abeaaf}

\bibitem[{{Ehrenreich} {et~al.}(2015){Ehrenreich}, {Bourrier}, {Wheatley},
  {Lecavelier des Etangs}, {H{\'e}brard}, {Udry}, {Bonfils}, {Delfosse},
  {D{\'e}sert}, {Sing}, \& {Vidal-Madjar}}]{2015Natur.522..459E}
{Ehrenreich}, D., {Bourrier}, V., {Wheatley}, P.~J., {et~al.} 2015, \nat, 522,
  459, \dodoi{10.1038/nature14501}

\bibitem[{{Feinstein} {et~al.}(2024){Feinstein}, {France}, {Cauley}, \&
  {Livingston}}]{2024RNAAS...8...86F}
{Feinstein}, A.~D., {France}, K., {Cauley}, P.~W., \& {Livingston}, J.~H. 2024,
  Research Notes of the American Astronomical Society, 8, 86,
  \dodoi{10.3847/2515-5172/ad35b7}

\bibitem[{{Feinstein} {et~al.}(2021){Feinstein}, {Montet}, {Johnson}, {Bean},
  {David}, {Gully-Santiago}, {Livingston}, \& {Luger}}]{feinstein21}
{Feinstein}, A.~D., {Montet}, B.~T., {Johnson}, M.~C., {et~al.} 2021, \aj, 162,
  213, \dodoi{10.3847/1538-3881/ac1f24}

\bibitem[{{Feinstein} {et~al.}(2022){Feinstein}, {France}, {Youngblood},
  {Duvvuri}, {Teal}, {Cauley}, {Seligman}, {Gaidos}, {Kempton}, {Bean},
  {Diamond-Lowe}, {Newton}, {Ginzburg}, {Plavchan}, {Gao}, \&
  {Schlichting}}]{2022AJ....164..110F}
{Feinstein}, A.~D., {France}, K., {Youngblood}, A., {et~al.} 2022, \aj, 164,
  110, \dodoi{10.3847/1538-3881/ac8107}

\bibitem[{{Fontenla} {et~al.}(2014){Fontenla}, {Landi}, {Snow}, \&
  {Woods}}]{2014SoPh..289..515F}
{Fontenla}, J.~M., {Landi}, E., {Snow}, M., \& {Woods}, T. 2014, \solphys, 289,
  515, \dodoi{10.1007/s11207-013-0431-4}

\bibitem[{{Foreman-Mackey} {et~al.}(2013){Foreman-Mackey}, {Hogg}, {Lang}, \&
  {Goodman}}]{2013PASP..125..306F}
{Foreman-Mackey}, D., {Hogg}, D.~W., {Lang}, D., \& {Goodman}, J. 2013,
  Publications of the Astronomical Society of the Pacific, 125, 306,
  \dodoi{10.1086/670067}

\bibitem[{{France} {et~al.}(2020){France}, {Duvvuri}, {Egan}, {Koskinen},
  {Wilson}, {Youngblood}, {Froning}, {Brown}, {Alvarado-G{\'o}mez},
  {Berta-Thompson}, {Drake}, {Garraffo}, {Kaltenegger}, {Kowalski}, {Linsky},
  {Loyd}, {Mauas}, {Miguel}, {Pineda}, {Rugheimer}, {Schneider}, {Tian}, \&
  {Vieytes}}]{2020AJ....160..237F}
{France}, K., {Duvvuri}, G., {Egan}, H., {et~al.} 2020, \aj, 160, 237,
  \dodoi{10.3847/1538-3881/abb465}

\bibitem[{{Fulton} {et~al.}(2017){Fulton}, {Petigura}, {Howard}, {Isaacson},
  {Marcy}, {Cargile}, {Hebb}, {Weiss}, {Johnson}, {Morton}, {Sinukoff},
  {Crossfield}, \& {Hirsch}}]{2017AJ....154..109F}
{Fulton}, B.~J., {Petigura}, E.~A., {Howard}, A.~W., {et~al.} 2017, \aj, 154,
  109, \dodoi{10.3847/1538-3881/aa80eb}

\bibitem[{{Gaia Collaboration} {et~al.}(2018){Gaia Collaboration}, {Brown},
  {Vallenari}, {Prusti}, {de Bruijne}, {Babusiaux}, {Bailer-Jones}, {Biermann},
  {Evans}, {Eyer}, {Jansen}, {Jordi}, {Klioner}, {Lammers}, {Lindegren},
  {Luri}, {Mignard}, {Panem}, {Pourbaix}, {Randich}, {Sartoretti}, {Siddiqui},
  {Soubiran}, {van Leeuwen}, {Walton}, {Arenou}, {Bastian}, {Cropper},
  {Drimmel}, {Katz}, {Lattanzi}, {Bakker}, {Cacciari}, {Casta{\~n}eda},
  {Chaoul}, {Cheek}, {De Angeli}, {Fabricius}, {Guerra}, {Holl}, {Masana},
  {Messineo}, {Mowlavi}, {Nienartowicz}, {Panuzzo}, {Portell}, {Riello},
  {Seabroke}, {Tanga}, {Th{\'e}venin}, {Gracia-Abril}, {Comoretto},
  {Garcia-Reinaldos}, {Teyssier}, {Altmann}, {Andrae}, {Audard},
  {Bellas-Velidis}, {Benson}, {Berthier}, {Blomme}, {Burgess}, {Busso},
  {Carry}, {Cellino}, {Clementini}, {Clotet}, {Creevey}, {Davidson}, {De
  Ridder}, {Delchambre}, {Dell'Oro}, {Ducourant},
  {Fern{\'a}ndez-Hern{\'a}ndez}, {Fouesneau}, {Fr{\'e}mat}, {Galluccio},
  {Garc{\'\i}a-Torres}, {Gonz{\'a}lez-N{\'u}{\~n}ez}, {Gonz{\'a}lez-Vidal},
  {Gosset}, {Guy}, {Halbwachs}, {Hambly}, {Harrison}, {Hern{\'a}ndez},
  {Hestroffer}, {Hodgkin}, {Hutton}, {Jasniewicz}, {Jean-Antoine-Piccolo},
  {Jordan}, {Korn}, {Krone-Martins}, {Lanzafame}, {Lebzelter}, {L{\"o}ffler},
  {Manteiga}, {Marrese}, {Mart{\'\i}n-Fleitas}, {Moitinho}, {Mora}, {Muinonen},
  {Osinde}, {Pancino}, {Pauwels}, {Petit}, {Recio-Blanco}, {Richards},
  {Rimoldini}, {Robin}, {Sarro}, {Siopis}, {Smith}, {Sozzetti}, {S{\"u}veges},
  {Torra}, {van Reeven}, {Abbas}, {Abreu Aramburu}, {Accart}, {Aerts},
  {Altavilla}, {{\'A}lvarez}, {Alvarez}, {Alves}, {Anderson}, {Andrei},
  {Anglada Varela}, {Antiche}, {Antoja}, {Arcay}, {Astraatmadja}, {Bach},
  {Baker}, {Balaguer-N{\'u}{\~n}ez}, {Balm}, {Barache}, {Barata}, {Barbato},
  {Barblan}, {Barklem}, {Barrado}, {Barros}, {Barstow}, {Bartholom{\'e}
  Mu{\~n}oz}, {Bassilana}, {Becciani}, {Bellazzini}, {Berihuete}, {Bertone},
  {Bianchi}, {Bienaym{\'e}}, {Blanco-Cuaresma}, {Boch}, {Boeche}, {Bombrun},
  {Borrachero}, {Bossini}, {Bouquillon}, {Bourda}, {Bragaglia}, {Bramante},
  {Breddels}, {Bressan}, {Brouillet}, {Br{\"u}semeister}, {Brugaletta},
  {Bucciarelli}, {Burlacu}, {Busonero}, {Butkevich}, {Buzzi}, {Caffau},
  {Cancelliere}, {Cannizzaro}, {Cantat-Gaudin}, {Carballo}, {Carlucci},
  {Carrasco}, {Casamiquela}, {Castellani}, {Castro-Ginard}, {Charlot},
  {Chemin}, {Chiavassa}, {Cocozza}, {Costigan}, {Cowell}, {Crifo}, {Crosta},
  {Crowley}, {Cuypers}, {Dafonte}, {Damerdji}, {Dapergolas}, {David}, {David},
  {de Laverny}, {De Luise}, {De March}, {de Martino}, {de Souza}, {de Torres},
  {Debosscher}, {del Pozo}, {Delbo}, {Delgado}, {Delgado}, {Di Matteo},
  {Diakite}, {Diener}, {Distefano}, {Dolding}, {Drazinos}, {Dur{\'a}n},
  {Edvardsson}, {Enke}, {Eriksson}, {Esquej}, {Eynard Bontemps}, {Fabre},
  {Fabrizio}, {Faigler}, {Falc{\~a}o}, {Farr{\`a}s Casas}, {Federici},
  {Fedorets}, {Fernique}, {Figueras}, {Filippi}, {Findeisen}, {Fonti},
  {Fraile}, {Fraser}, {Fr{\'e}zouls}, {Gai}, {Galleti}, {Garabato},
  {Garc{\'\i}a-Sedano}, {Garofalo}, {Garralda}, {Gavel}, {Gavras}, {Gerssen},
  {Geyer}, {Giacobbe}, {Gilmore}, {Girona}, {Giuffrida}, {Glass}, {Gomes},
  {Granvik}, {Gueguen}, {Guerrier}, {Guiraud}, {Guti{\'e}rrez-S{\'a}nchez},
  {Haigron}, {Hatzidimitriou}, {Hauser}, {Haywood}, {Heiter}, {Helmi}, {Heu},
  {Hilger}, {Hobbs}, {Hofmann}, {Holland}, {Huckle}, {Hypki}, {Icardi},
  {Jan{\ss}en}, {Jevardat de Fombelle}, {Jonker}, {Juh{\'a}sz}, {Julbe},
  {Karampelas}, {Kewley}, {Klar}, {Kochoska}, {Kohley}, {Kolenberg},
  {Kontizas}, {Kontizas}, {Koposov}, {Kordopatis}, {Kostrzewa-Rutkowska},
  {Koubsky}, {Lambert}, {Lanza}, {Lasne}, {Lavigne}, {Le Fustec}, {Le
  Poncin-Lafitte}, {Lebreton}, {Leccia}, {Leclerc}, {Lecoeur-Taibi},
  {Lenhardt}, {Leroux}, {Liao}, {Licata}, {Lindstr{\o}m}, {Lister}, {Livanou},
  {Lobel}, {L{\'o}pez}, {Managau}, {Mann}, {Mantelet}, {Marchal}, {Marchant},
  {Marconi}, {Marinoni}, {Marschalk{\'o}}, {Marshall}, {Martino}, {Marton},
  {Mary}, {Massari}, {Matijevi{\v{c}}}, {Mazeh}, {McMillan}, {Messina},
  {Michalik}, {Millar}, {Molina}, {Molinaro}, {Moln{\'a}r}, {Montegriffo},
  {Mor}, {Morbidelli}, {Morel}, {Morris}, {Mulone}, {Muraveva}, {Musella},
  {Nelemans}, {Nicastro}, {Noval}, {O'Mullane}, {Ord{\'e}novic},
  {Ord{\'o}{\~n}ez-Blanco}, {Osborne}, {Pagani}, {Pagano}, {Pailler},
  {Palacin}, {Palaversa}, {Panahi}, {Pawlak}, {Piersimoni}, {Pineau}, {Plachy},
  {Plum}, {Poggio}, {Poujoulet}, {Pr{\v{s}}a}, {Pulone}, {Racero}, {Ragaini},
  {Rambaux}, {Ramos-Lerate}, {Regibo}, {Reyl{\'e}}, {Riclet}, {Ripepi}, {Riva},
  {Rivard}, {Rixon}, {Roegiers}, {Roelens}, {Romero-G{\'o}mez}, {Rowell},
  {Royer}, {Ruiz-Dern}, {Sadowski}, {Sagrist{\`a} Sell{\'e}s}, {Sahlmann},
  {Salgado}, {Salguero}, {Sanna}, {Santana-Ros}, {Sarasso}, {Savietto},
  {Schultheis}, {Sciacca}, {Segol}, {Segovia}, {S{\'e}gransan}, {Shih},
  {Siltala}, {Silva}, {Smart}, {Smith}, {Solano}, {Solitro}, {Sordo}, {Soria
  Nieto}, {Souchay}, {Spagna}, {Spoto}, {Stampa}, {Steele},
  {Steidelm{\"u}ller}, {Stephenson}, {Stoev}, {Suess}, {Surdej}, {Szabados},
  {Szegedi-Elek}, {Tapiador}, {Taris}, {Tauran}, {Taylor}, {Teixeira},
  {Terrett}, {Teyssandier}, {Thuillot}, {Titarenko}, {Torra Clotet}, {Turon},
  {Ulla}, {Utrilla}, {Uzzi}, {Vaillant}, {Valentini}, {Valette}, {van Elteren},
  {Van Hemelryck}, {van Leeuwen}, {Vaschetto}, {Vecchiato}, {Veljanoski},
  {Viala}, {Vicente}, {Vogt}, {von Essen}, {Voss}, {Votruba}, {Voutsinas},
  {Walmsley}, {Weiler}, {Wertz}, {Wevers}, {Wyrzykowski}, {Yoldas},
  {{\v{Z}}erjal}, {Ziaeepour}, {Zorec}, {Zschocke}, {Zucker}, {Zurbach}, \&
  {Zwitter}}]{2018AA...616A...1G}
{Gaia Collaboration}, {Brown}, A.~G.~A., {Vallenari}, A., {et~al.} 2018, \aap,
  616, A1, \dodoi{10.1051/0004-6361/201833051}

\bibitem[{{Garc{\'\i}a Mu{\~n}oz} {et~al.}(2021){Garc{\'\i}a Mu{\~n}oz},
  {Fossati}, {Youngblood}, {Nettelmann}, {Gandolfi}, {Cabrera}, \&
  {Rauer}}]{2021ApJ...907L..36G}
{Garc{\'\i}a Mu{\~n}oz}, A., {Fossati}, L., {Youngblood}, A., {et~al.} 2021,
  \apjl, 907, L36, \dodoi{10.3847/2041-8213/abd9b8}

\bibitem[{{Gilbert} {et~al.}(2022){Gilbert}, {Barclay}, {Quintana},
  {Walkowicz}, {Vega}, {Schlieder}, {Monsue}, {Cale}, {Collins}, {Gaidos}, {El
  Mufti}, {Reefe}, {Plavchan}, {Tanner}, {Wittenmyer}, {Wittrock}, {Jenkins},
  {Latham}, {Ricker}, {Rose}, {Seager}, {Vanderspek}, \&
  {Winn}}]{2022AJ....163..147G}
{Gilbert}, E.~A., {Barclay}, T., {Quintana}, E.~V., {et~al.} 2022, \aj, 163,
  147, \dodoi{10.3847/1538-3881/ac23ca}

\bibitem[{{Ginzburg} {et~al.}(2018){Ginzburg}, {Schlichting}, \&
  {Sari}}]{2018MNRAS.476..759G}
{Ginzburg}, S., {Schlichting}, H.~E., \& {Sari}, R. 2018, \mnras, 476, 759,
  \dodoi{10.1093/mnras/sty290}

\bibitem[{Harris {et~al.}(2020)Harris, Millman, van~der Walt, Gommers,
  Virtanen, Cournapeau, Wieser, Taylor, Berg, Smith, Kern, Picus, Hoyer, van
  Kerkwijk, Brett, Haldane, del R{'{\i}}o, Wiebe, Peterson,
  G{'{e}}rard-Marchant, Sheppard, Reddy, Weckesser, Abbasi, Gohlke, \&
  Oliphant}]{harris2020array}
Harris, C.~R., Millman, K.~J., van~der Walt, S.~J., {et~al.} 2020, Nature, 585,
  357, \dodoi{10.1038/s41586-020-2649-2}

\bibitem[{{Hirano} {et~al.}(2020){Hirano}, {Krishnamurthy}, {Gaidos},
  {Flewelling}, {Mann}, {Narita}, {Plavchan}, {Kotani}, {Tamura}, {Harakawa},
  {Hodapp}, {Ishizuka}, {Jacobson}, {Konishi}, {Kudo}, {Kurokawa}, {Kuzuhara},
  {Nishikawa}, {Omiya}, {Serizawa}, {Ueda}, \& {Vievard}}]{hirano20}
{Hirano}, T., {Krishnamurthy}, V., {Gaidos}, E., {et~al.} 2020, \apjl, 899,
  L13, \dodoi{10.3847/2041-8213/aba6eb}

\bibitem[{{Howard} \& {MacGregor}(2022)}]{2022ApJ...926..204H}
{Howard}, W.~S., \& {MacGregor}, M.~A. 2022, \apj, 926, 204,
  \dodoi{10.3847/1538-4357/ac426e}

\bibitem[{Hunter(2007)}]{hunter2007matplotlib}
Hunter, J.~D. 2007, Computing in science \& engineering, 9, 90

\bibitem[{{Jin} {et~al.}(2014){Jin}, {Mordasini}, {Parmentier}, {van Boekel},
  {Henning}, \& {Ji}}]{2014ApJ...795...65J}
{Jin}, S., {Mordasini}, C., {Parmentier}, V., {et~al.} 2014, \apj, 795, 65,
  \dodoi{10.1088/0004-637X/795/1/65}

\bibitem[{{Khodachenko} {et~al.}(2019){Khodachenko}, {Shaikhislamov}, {Lammer},
  {Berezutsky}, {Miroshnichenko}, {Rumenskikh}, {Kislyakova}, \&
  {Dwivedi}}]{2019ApJ...885...67K}
{Khodachenko}, M.~L., {Shaikhislamov}, I.~F., {Lammer}, H., {et~al.} 2019,
  \apj, 885, 67, \dodoi{10.3847/1538-4357/ab46a4}

\bibitem[{{King} {et~al.}(2018){King}, {Wheatley}, {Salz}, {Bourrier},
  {Czesla}, {Ehrenreich}, {Kirk}, {Lecavelier des Etangs}, {Louden}, {Schmitt},
  \& {Schneider}}]{2018MNRAS.478.1193K}
{King}, G.~W., {Wheatley}, P.~J., {Salz}, M., {et~al.} 2018, \mnras, 478, 1193,
  \dodoi{10.1093/mnras/sty1110}

\bibitem[{{Klein} {et~al.}(2021){Klein}, {Donati}, {Moutou}, {Delfosse},
  {Bonfils}, {Martioli}, {Fouqu{\'e}}, {Cloutier}, {Artigau}, {Doyon},
  {H{\'e}brard}, {Morin}, {Rameau}, {Plavchan}, \&
  {Gaidos}}]{2021MNRAS.502..188K}
{Klein}, B., {Donati}, J.-F., {Moutou}, C., {et~al.} 2021, \mnras, 502, 188,
  \dodoi{10.1093/mnras/staa3702}

\bibitem[{{Kreidberg}(2015)}]{2015PASP..127.1161K}
{Kreidberg}, L. 2015, \pasp, 127, 1161, \dodoi{10.1086/683602}

\bibitem[{{Kulow} {et~al.}(2014){Kulow}, {France}, {Linsky}, \&
  {Loyd}}]{2014ApJ...786..132K}
{Kulow}, J.~R., {France}, K., {Linsky}, J., \& {Loyd}, R.~O.~P. 2014, \apj,
  786, 132, \dodoi{10.1088/0004-637X/786/2/132}

\bibitem[{{Lannier} {et~al.}(2017){Lannier}, {Lagrange}, {Bonavita},
  {Borgniet}, {Delorme}, {Meunier}, {Desidera}, {Messina}, {Chauvin}, \&
  {Keppler}}]{2017AA...603A..54L}
{Lannier}, J., {Lagrange}, A.~M., {Bonavita}, M., {et~al.} 2017, \aap, 603,
  A54, \dodoi{10.1051/0004-6361/201628677}

\bibitem[{{Lavie} {et~al.}(2017){Lavie}, {Ehrenreich}, {Bourrier}, {Lecavelier
  des Etangs}, {Vidal-Madjar}, {Delfosse}, {Gracia Berna}, {Heng}, {Thomas},
  {Udry}, \& {Wheatley}}]{2017A&A...605L...7L}
{Lavie}, B., {Ehrenreich}, D., {Bourrier}, V., {et~al.} 2017, \aap, 605, L7,
  \dodoi{10.1051/0004-6361/201731340}

\bibitem[{{Lee} \& {Connors}(2021)}]{2021ApJ...908...32L}
{Lee}, E.~J., \& {Connors}, N.~J. 2021, \apj, 908, 32,
  \dodoi{10.3847/1538-4357/abd6c7}

\bibitem[{{Leto} {et~al.}(2000){Leto}, {Pagano}, {Linsky}, {Rodon{\`o}}, \&
  {Umana}}]{2000A&A...359.1035L}
{Leto}, G., {Pagano}, I., {Linsky}, J.~L., {Rodon{\`o}}, M., \& {Umana}, G.
  2000, \aap, 359, 1035

\bibitem[{{Lightkurve Collaboration} {et~al.}(2018){Lightkurve Collaboration},
  {Cardoso}, {Hedges}, {Gully-Santiago}, {Saunders}, {Cody}, {Barclay}, {Hall},
  {Sagear}, {Turtelboom}, {Zhang}, {Tzanidakis}, {Mighell}, {Coughlin}, {Bell},
  {Berta-Thompson}, {Williams}, {Dotson}, \& {Barentsen}}]{2018ascl.soft12013L}
{Lightkurve Collaboration}, {Cardoso}, J.~V.~d.~M., {Hedges}, C., {et~al.}
  2018, {Lightkurve: Kepler and TESS time series analysis in Python},
  Astrophysics Source Code Library.
\newblock \doeprint{1812.013}

\bibitem[{{Linsky} {et~al.}(2006){Linsky}, {Draine}, {Moos}, {Jenkins}, {Wood},
  {Oliveira}, {Blair}, {Friedman}, {Gry}, {Knauth}, {Kruk}, {Lacour}, {Lehner},
  {Redfield}, {Shull}, {Sonneborn}, \& {Williger}}]{2006ApJ...647.1106L}
{Linsky}, J.~L., {Draine}, B.~T., {Moos}, H.~W., {et~al.} 2006, \apj, 647,
  1106, \dodoi{10.1086/505556}

\bibitem[{{Lopez} \& {Fortney}(2013)}]{2013ApJ...776....2L}
{Lopez}, E.~D., \& {Fortney}, J.~J. 2013, \apj, 776, 2,
  \dodoi{10.1088/0004-637X/776/1/2}

\bibitem[{{Loyd} \& {France}(2014)}]{2014ApJS..211....9L}
{Loyd}, R.~O.~P., \& {France}, K. 2014, \apjs, 211, 9,
  \dodoi{10.1088/0067-0049/211/1/9}

\bibitem[{{Loyd} {et~al.}(2018){Loyd}, {France}, {Youngblood}, {Schneider},
  {Brown}, {Hu}, {Segura}, {Linsky}, {Redfield}, {Tian}, {Rugheimer}, {Miguel},
  \& {Froning}}]{2018ApJ...867...71L}
{Loyd}, R.~O.~P., {France}, K., {Youngblood}, A., {et~al.} 2018, \apj, 867, 71,
  \dodoi{10.3847/1538-4357/aae2bd}

\bibitem[{{Lundkvist} {et~al.}(2016){Lundkvist}, {Kjeldsen}, {Albrecht},
  {Davies}, {Basu}, {Huber}, {Justesen}, {Karoff}, {Silva Aguirre}, {van
  Eylen}, {Vang}, {Arentoft}, {Barclay}, {Bedding}, {Campante}, {Chaplin},
  {Christensen-Dalsgaard}, {Elsworth}, {Gilliland}, {Handberg}, {Hekker},
  {Kawaler}, {Lund}, {Metcalfe}, {Miglio}, {Rowe}, {Stello}, {Tingley}, \&
  {White}}]{2016NatCo...711201L}
{Lundkvist}, M.~S., {Kjeldsen}, H., {Albrecht}, S., {et~al.} 2016, Nature
  Communications, 7, 11201, \dodoi{10.1038/ncomms11201}

\bibitem[{{Mallorqu{\'\i}n} {et~al.}(2024){Mallorqu{\'\i}n}, {B{\'e}jar},
  {Lodieu}, {Zapatero Osorio}, {Yu}, {Su{\'a}rez Mascare{\~n}o}, {Damasso},
  {Sanz-Forcada}, {Ribas}, {Reiners}, {Quirrenbach}, {Amado}, {Caballero},
  {Aigrain}, {Barrag{\'a}n}, {Dreizler}, {Fern{\'a}ndez-Mart{\'\i}n}, {Goffo},
  {Henning}, {Kaminski}, {Klein}, {Luque}, {Montes}, {Morales}, {Nagel},
  {Pall{\'e}}, {Reffert}, {Schlecker}, \& {Schweitzer}}]{2024AA...689A.132M}
{Mallorqu{\'\i}n}, M., {B{\'e}jar}, V.~J.~S., {Lodieu}, N., {et~al.} 2024,
  \aap, 689, A132, \dodoi{10.1051/0004-6361/202450047}

\bibitem[{{Mamajek} \& {Bell}(2014)}]{2014MNRAS.445.2169M}
{Mamajek}, E.~E., \& {Bell}, C. P.~M. 2014, \mnras, 445, 2169,
  \dodoi{10.1093/mnras/stu1894}

\bibitem[{{Martioli} {et~al.}(2021){Martioli}, {H{\'e}brard}, {Correia},
  {Laskar}, \& {Lecavelier des Etangs}}]{2021AA...649A.177M}
{Martioli}, E., {H{\'e}brard}, G., {Correia}, A.~C.~M., {Laskar}, J., \&
  {Lecavelier des Etangs}, A. 2021, \aap, 649, A177,
  \dodoi{10.1051/0004-6361/202040235}

\bibitem[{{McCann} {et~al.}(2019){McCann}, {Murray-Clay}, {Kratter}, \&
  {Krumholz}}]{2019ApJ...873...89M}
{McCann}, J., {Murray-Clay}, R.~A., {Kratter}, K., \& {Krumholz}, M.~R. 2019,
  \apj, 873, 89, \dodoi{10.3847/1538-4357/ab05b8}

\bibitem[{{Murray-Clay} {et~al.}(2009){Murray-Clay}, {Chiang}, \&
  {Murray}}]{2009ApJ...693...23M}
{Murray-Clay}, R.~A., {Chiang}, E.~I., \& {Murray}, N. 2009, \apj, 693, 23,
  \dodoi{10.1088/0004-637X/693/1/23}

\bibitem[{{Oklop{\v{c}}i{\'c}} \& {Hirata}(2018)}]{2018ApJ...855L..11O}
{Oklop{\v{c}}i{\'c}}, A., \& {Hirata}, C.~M. 2018, \apjl, 855, L11,
  \dodoi{10.3847/2041-8213/aaada9}

\bibitem[{{Owen} \& {Alvarez}(2016)}]{2016ApJ...816...34O}
{Owen}, J.~E., \& {Alvarez}, M.~A. 2016, \apj, 816, 34,
  \dodoi{10.3847/0004-637X/816/1/34}

\bibitem[{{Owen} \& {Lai}(2018)}]{2018MNRAS.479.5012O}
{Owen}, J.~E., \& {Lai}, D. 2018, \mnras, 479, 5012,
  \dodoi{10.1093/mnras/sty1760}

\bibitem[{{Owen} \& {Wu}(2017)}]{2017ApJ...847...29O}
{Owen}, J.~E., \& {Wu}, Y. 2017, \apj, 847, 29,
  \dodoi{10.3847/1538-4357/aa890a}

\bibitem[{{Owen} {et~al.}(2023){Owen}, {Murray-Clay}, {Schreyer},
  {Schlichting}, {Ardila}, {Gupta}, {Loyd}, {Shkolnik}, {Sing}, \&
  {Swain}}]{2023MNRAS.518.4357O}
{Owen}, J.~E., {Murray-Clay}, R.~A., {Schreyer}, E., {et~al.} 2023, \mnras,
  518, 4357, \dodoi{10.1093/mnras/stac3414}

\bibitem[{{Pagano} {et~al.}(2000){Pagano}, {Linsky}, {Carkner}, {Robinson},
  {Woodgate}, \& {Timothy}}]{2000ApJ...532..497P}
{Pagano}, I., {Linsky}, J.~L., {Carkner}, L., {et~al.} 2000, \apj, 532, 497,
  \dodoi{10.1086/308559}

\bibitem[{{Pallavicini} {et~al.}(1990){Pallavicini}, {Tagliaferri}, \&
  {Stella}}]{1990A&A...228..403P}
{Pallavicini}, R., {Tagliaferri}, G., \& {Stella}, L. 1990, \aap, 228, 403

\bibitem[{{Peacock} {et~al.}(2020){Peacock}, {Barman}, {Shkolnik}, {Loyd},
  {Schneider}, {Pagano}, \& {Meadows}}]{2020ApJ...895....5P}
{Peacock}, S., {Barman}, T., {Shkolnik}, E.~L., {et~al.} 2020, \apj, 895, 5,
  \dodoi{10.3847/1538-4357/ab893a}

\bibitem[{{Plavchan} {et~al.}(2009){Plavchan}, {Werner}, {Chen}, {Stapelfeldt},
  {Su}, {Stauffer}, \& {Song}}]{2009ApJ...698.1068P}
{Plavchan}, P., {Werner}, M.~W., {Chen}, C.~H., {et~al.} 2009, \apj, 698, 1068,
  \dodoi{10.1088/0004-637X/698/2/1068}

\bibitem[{{Plavchan} {et~al.}(2020){Plavchan}, {Barclay}, {Gagn{\'e}}, {Gao},
  {Cale}, {Matzko}, {Dragomir}, {Quinn}, {Feliz}, {Stassun}, {Crossfield},
  {Berardo}, {Latham}, {Tieu}, {Anglada-Escud{\'e}}, {Ricker}, {Vanderspek},
  {Seager}, {Winn}, {Jenkins}, {Rinehart}, {Krishnamurthy}, {Dynes}, {Doty},
  {Adams}, {Afanasev}, {Beichman}, {Bottom}, {Bowler}, {Brinkworth}, {Brown},
  {Cancino}, {Ciardi}, {Clampin}, {Clark}, {Collins}, {Davison},
  {Foreman-Mackey}, {Furlan}, {Gaidos}, {Geneser}, {Giddens}, {Gilbert},
  {Hall}, {Hellier}, {Henry}, {Horner}, {Howard}, {Huang}, {Huber}, {Kane},
  {Kenworthy}, {Kielkopf}, {Kipping}, {Klenke}, {Kruse}, {Latouf}, {Lowrance},
  {Mennesson}, {Mengel}, {Mills}, {Morton}, {Narita}, {Newton}, {Nishimoto},
  {Okumura}, {Palle}, {Pepper}, {Quintana}, {Roberge}, {Roccatagliata},
  {Schlieder}, {Tanner}, {Teske}, {Tinney}, {Vanderburg}, {von Braun}, {Walp},
  {Wang}, {Wang}, {Weigand}, {White}, {Wittenmyer}, {Wright}, {Youngblood},
  {Zhang}, \& {Zilberman}}]{2020Natur.582..497P}
{Plavchan}, P., {Barclay}, T., {Gagn{\'e}}, J., {et~al.} 2020, \nat, 582, 497,
  \dodoi{10.1038/s41586-020-2400-z}

\bibitem[{{Rockcliffe} {et~al.}(2023){Rockcliffe}, {Newton}, {Youngblood},
  {Duvvuri}, {Plavchan}, {Gao}, {Mann}, \& {Lowrance}}]{2023AJ....166...77R}
{Rockcliffe}, K.~E., {Newton}, E.~R., {Youngblood}, A., {et~al.} 2023, \aj,
  166, 77, \dodoi{10.3847/1538-3881/ace536}

\bibitem[{{Rockcliffe} {et~al.}(2021){Rockcliffe}, {Newton}, {Youngblood},
  {Bourrier}, {Mann}, {Berta-Thompson}, {Ag{\"u}eros}, {N{\'u}{\~n}ez}, \&
  {Charbonneau}}]{2021AJ....162..116R}
---. 2021, \aj, 162, 116, \dodoi{10.3847/1538-3881/ac126f}

\bibitem[{{Sanz-Forcada} {et~al.}(2011){Sanz-Forcada}, {Micela}, {Ribas},
  {Pollock}, {Eiroa}, {Velasco}, {Solano}, \&
  {Garc{\'\i}a-{\'A}lvarez}}]{2011A&A...532A...6S}
{Sanz-Forcada}, J., {Micela}, G., {Ribas}, I., {et~al.} 2011, \aap, 532, A6,
  \dodoi{10.1051/0004-6361/201116594}

\bibitem[{{Schlawin} {et~al.}(2021){Schlawin}, {Ilyin}, {Feinstein}, {Bean},
  {Huang}, {Gao}, {Strassmeier}, \& {Poppenhaeger}}]{schlawin21}
{Schlawin}, E., {Ilyin}, I., {Feinstein}, A.~D., {et~al.} 2021, Research Notes
  of the American Astronomical Society, 5, 195,
  \dodoi{10.3847/2515-5172/ac1f2f}

\bibitem[{{Sing} {et~al.}(2019){Sing}, {Lavvas}, {Ballester}, {Lecavelier des
  Etangs}, {Marley}, {Nikolov}, {Ben-Jaffel}, {Bourrier}, {Buchhave}, {Deming},
  {Ehrenreich}, {Mikal-Evans}, {Kataria}, {Lewis}, {L{\'o}pez-Morales},
  {Garc{\'\i}a Mu{\~n}oz}, {Henry}, {Sanz-Forcada}, {Spake}, {Wakeford}, \&
  {PanCET Collaboration}}]{2019AJ....158...91S}
{Sing}, D.~K., {Lavvas}, P., {Ballester}, G.~E., {et~al.} 2019, \aj, 158, 91,
  \dodoi{10.3847/1538-3881/ab2986}

\bibitem[{{Stef{\`a}nsson} {et~al.}(2022){Stef{\`a}nsson}, {Mahadevan},
  {Petrovich}, {Winn}, {Kanodia}, {Millholland}, {Maney}, {Ca{\~n}as},
  {Wisniewski}, {Robertson}, {Ninan}, {Ford}, {Bender}, {Blake}, {Cegla},
  {Cochran}, {Diddams}, {Dong}, {Endl}, {Fredrick}, {Halverson}, {Hearty},
  {Hebb}, {Hirano}, {Lin}, {Logsdon}, {Lubar}, {McElwain}, {Metcalf}, {Monson},
  {Rajagopal}, {Ramsey}, {Roy}, {Schwab}, {Schweiker}, {Terrien}, \&
  {Wright}}]{2022ApJ...931L..15S}
{Stef{\`a}nsson}, G., {Mahadevan}, S., {Petrovich}, C., {et~al.} 2022, \apjl,
  931, L15, \dodoi{10.3847/2041-8213/ac6e3c}

\bibitem[{{Tilipman} {et~al.}(2021){Tilipman}, {Vieytes}, {Linsky}, {Buccino},
  \& {France}}]{2021ApJ...909...61T}
{Tilipman}, D., {Vieytes}, M., {Linsky}, J.~L., {Buccino}, A.~P., \& {France},
  K. 2021, \apj, 909, 61, \dodoi{10.3847/1538-4357/abd62f}

\bibitem[{{Tristan} {et~al.}(2023){Tristan}, {Notsu}, {Kowalski}, {Brown},
  {Wisniewski}, {Osten}, {Vrijmoet}, {White}, {Carter}, {Grady}, {Henry},
  {Hinojosa}, {Lomax}, {Neff}, {Paredes}, \& {Soutter}}]{2023ApJ...951...33T}
{Tristan}, I.~I., {Notsu}, Y., {Kowalski}, A.~F., {et~al.} 2023, \apj, 951, 33,
  \dodoi{10.3847/1538-4357/acc94f}

\bibitem[{{Turnbull}(2015)}]{2015arXiv151001731T}
{Turnbull}, M.~C. 2015, arXiv e-prints, arXiv:1510.01731,
  \dodoi{10.48550/arXiv.1510.01731}

\bibitem[{{Vidal-Madjar} {et~al.}(2003){Vidal-Madjar}, {Lecavelier des Etangs},
  {D{\'e}sert}, {Ballester}, {Ferlet}, {H{\'e}brard}, \&
  {Mayor}}]{2003Natur.422..143V}
{Vidal-Madjar}, A., {Lecavelier des Etangs}, A., {D{\'e}sert}, J.~M., {et~al.}
  2003, \nat, 422, 143, \dodoi{10.1038/nature01448}

\bibitem[{Virtanen {et~al.}(2020)Virtanen, Gommers, Oliphant, Haberland, Reddy,
  Cournapeau, Burovski, Peterson, {Weckesser}, {Bright}, {van der Walt},
  {Brett}, {Wilson}, {Jarrod Millman}, {Mayorov}, {Nelson}, {Jones}, {Kern},
  {Larson}, {Carey}, {Polat}, {Feng}, {Moore}, {Vand erPlas}, {Laxalde},
  {Perktold}, {Cimrman}, {Henriksen}, {Quintero}, {Harris}, {Archibald},
  {Ribeiro}, {Pedregosa}, {van Mulbregt}, \& {Contributors}}]{2020SciPy}
Virtanen, P., Gommers, R., Oliphant, T.~E., {et~al.} 2020, Nature Methods

\bibitem[{{Wittrock} {et~al.}(2023){Wittrock}, {Plavchan}, {Cale}, {Barclay},
  {Ludwig}, {Schwarz}, {M{\'e}karnia}, {Triaud}, {Abe}, {Suarez}, {Guillot},
  {Conti}, {Collins}, {Waite}, {Kielkopf}, {Collins}, {Dreizler}, {El Mufti},
  {Feliz}, {Gaidos}, {Geneser}, {Horne}, {Kane}, {Lowrance}, {Martioli},
  {Radford}, {Reefe}, {Roccatagliata}, {Shporer}, {Stassun}, {Stockdale},
  {Tan}, {Tanner}, \& {Vega}}]{2023AJ....166..232W}
{Wittrock}, J.~M., {Plavchan}, P.~P., {Cale}, B.~L., {et~al.} 2023, \aj, 166,
  232, \dodoi{10.3847/1538-3881/acfda8}

\bibitem[{{Woodgate} {et~al.}(1998){Woodgate}, {Kimble}, {Bowers}, {Kraemer},
  {Kaiser}, {Danks}, {Grady}, {Loiacono}, {Brumfield}, {Feinberg}, {Gull},
  {Heap}, {Maran}, {Lindler}, {Hood}, {Meyer}, {Vanhouten}, {Argabright},
  {Franka}, {Bybee}, {Dorn}, {Bottema}, {Woodruff}, {Michika}, {Sullivan},
  {Hetlinger}, {Ludtke}, {Stocker}, {Delamere}, {Rose}, {Becker}, {Garner},
  {Timothy}, {Blouke}, {Joseph}, {Hartig}, {Green}, {Jenkins}, {Linsky},
  {Hutchings}, {Moos}, {Boggess}, {Roesler}, \&
  {Weistrop}}]{1998PASP..110.1183W}
{Woodgate}, B.~E., {Kimble}, R.~A., {Bowers}, C.~W., {et~al.} 1998, \pasp, 110,
  1183, \dodoi{10.1086/316243}

\bibitem[{{Youngblood} \& {Newton}(2022)}]{2022zndo...6949067Y}
{Youngblood}, A., \& {Newton}, E.~R. 2022, {allisony/lyapy: First release
  created for citation purposes in the literature}, v1.0.0, Zenodo,  Zenodo,
  \dodoi{10.5281/zenodo.6949067}

\bibitem[{{Youngblood} {et~al.}(2016){Youngblood}, {France}, {Loyd}, {Linsky},
  {Redfield}, {Schneider}, {Wood}, {Brown}, {Froning}, {Miguel}, {Rugheimer},
  \& {Walkowicz}}]{2016ApJ...824..101Y}
{Youngblood}, A., {France}, K., {Loyd}, R.~O.~P., {et~al.} 2016, \apj, 824,
  101, \dodoi{10.3847/0004-637X/824/2/101}

\bibitem[{{Zhang} {et~al.}(2022){Zhang}, {Knutson}, {Wang}, {Dai}, {dos
  Santos}, {Fossati}, {Henry}, {Ehrenreich}, {Alibert}, {Hoyer}, {Wilson}, \&
  {Bonfanti}}]{2022AJ....163...68Z}
{Zhang}, M., {Knutson}, H.~A., {Wang}, L., {et~al.} 2022, \aj, 163, 68,
  \dodoi{10.3847/1538-3881/ac3f3b}

\bibitem[{{Zicher} {et~al.}(2022){Zicher}, {Barrag{\'a}n}, {Klein}, {Aigrain},
  {Owen}, {Gandolfi}, {Lagrange}, {Serrano}, {Kaye}, {Nielsen}, {Rajpaul},
  {Grandjean}, {Goffo}, \& {Nicholson}}]{2022MNRAS.512.3060Z}
{Zicher}, N., {Barrag{\'a}n}, O., {Klein}, B., {et~al.} 2022, \mnras, 512,
  3060, \dodoi{10.1093/mnras/stac614}

\bibitem[{{Zuckerman} \& {Song}(2004)}]{2004ARA&A..42..685Z}
{Zuckerman}, B., \& {Song}, I. 2004, \araa, 42, 685,
  \dodoi{10.1146/annurev.astro.42.053102.134111}

\end{thebibliography}
\end{document}